\documentclass[conference, 10pt]{IEEEtran}

\usepackage{blindtext}

\usepackage{xpatch}

\usepackage{amsmath}
\usepackage{amssymb}
\usepackage{mathtools}
\usepackage{wasysym}
\usepackage{hyperref}
\usepackage{tikz}
\usepackage{listings}
\usepackage{multirow}
\usepackage{shuffle}
\usepackage{stackengine}
\usepackage{mathpartir}
\usepackage{amsthm}

\usetikzlibrary{arrows,automata,shapes.gates.logic.US,shapes.gates.logic.IEC,calc}

\usepackage[ruled,vlined]{algorithm2e}
\usepackage{subfiles}
\usepackage{hhline}
\usepackage[clock]{ifsym}
\usepackage{xspace}
\usepackage{wrapfig}
\usepackage{booktabs}
\usepackage{subcaption}
\usepackage{extarrows}
\usepackage{subfiles}

\usepackage{bbm}
\usepackage{fancyvrb}
\usepackage{mathpartir}

\usepackage[colorinlistoftodos,prependcaption]{todonotes}
\usepackage{xargs}
\usepackage{enumitem}
\usepackage{turnstile}
\usepackage{colortbl}
\usepackage{empheq}
\usepackage[most]{tcolorbox}
\usepackage{lipsum,adjustbox}
\usepackage{stmaryrd}
\usepackage[switch]{lineno}
\usepackage{etoolbox}

\newtoggle{inclComments}
\toggletrue{inclComments}

\newcommand{\uspecSub}{$\mu$specRE\xspace}

\newtcbox{\mycode}[1][]{%
nobeforeafter, math upper, tcbox raise base,
enhanced, colframe=black,
colback=gray!5, boxrule=0.5pt,
#1}

\definecolor{auburn}{rgb}{0.43, 0.21, 0.1}
\definecolor{cadmiumgreen}{rgb}{0.0, 0.42, 0.24}

\newcommand{\sinkv}{\mathsf{sink}}
\newcommand{\sourcev}{\mathsf{source}}

\newcommand{\nat}{\mathbb{N}}

\newcommand{\domain}{\textsf{Dom}}

\newcommand{\anevent}{\textsf{msg}}

\newcommand{\bnf}{\;\mid\;}

\newcommand{\init}{\mathsf{init}}

\newcommandx{\adwait}[2][1=]{\todo[inline, linecolor=red,backgroundcolor=purple!5,bordercolor=red,#1]{{\color{red}A:} #2}}

\stackMath

\newcommand{\rangeOf}[1]{\mathit{range}(#1)}

\newcommand{\final}{\mathsf{final}}
\newcommand{\rej}{\mathsf{rej}}

\newcommand{\axiomautoof}[1]{\mathsf{aa}(#1)}

\newcommand{\hbrel}{{\color{blue} \mathsf{hb}}}
\newcommand{\hbarrow}{\xrightarrow{\hbrel}}
\newcommand{\muspec}{$\mu$spec}

\newcommand{\aprogram}{\mathcal{P}}
\newcommand{\istream}{\mathcal{I}}
\newcommand{\eventsOf}[1]{\mathsf{eventsOf}(#1)}

\newcommand{\aformula}{\phi}

\newcommand{\anaxiom}{\textsc{ax}}
\newcommand{\acontext}{\mathsf{cxt}}

\newcommand{\setstages}{\mathsf{Stages}}
\newcommand{\setoperations}{\mathbb{O}}
\newcommand{\setinstr}{\mathbb{I}}

\newcommand{\setcores}{\mathsf{Cores}}
\newcommand{\setevents}{\mathbb{E}}
\newcommand{\eventset}{\mathsf{E}}
\newcommand{\astage}{\mathsf{st}}
\newcommand{\asymbinstr}{\mathtt{i}}
\newcommand{\symbinstrset}{\mathtt{I}}
\newcommand{\aninstr}{\mathsf{i}}
\newcommand{\anoperation}{\mathsf{o}}

\newcommand{\acore}{\mathtt{c}}

\newcommand{\fetchstage}{\mathtt{Fet}}
\newcommand{\decodestage}{\mathtt{Dec}}
\newcommand{\execstage}{\mathtt{Exe}}
\newcommand{\wbstage}{\mathtt{WB}}

\newcommand{\aload}{\texttt{load}}
\newcommand{\astore}{\texttt{store}}
\newcommand{\anadd}{\texttt{add}}
\newcommand{\amov}{\texttt{mov}}

\newcommand{\destof}[1]{\texttt{dest}(#1)}
\newcommand{\srconeof}[1]{\texttt{src1}(#1)}
\newcommand{\srctwoof}[1]{\texttt{src2}(#1)}

\newcommand{\atrace}{\sigma}
\newcommand{\prefixClosureOf}[1]{\mathsf{pfxClos}(#1)}

\renewcommand{\anevent}{\mathsf{e}}
\newcommand{\event}[2]{#1.#2}

\newcommand{\anatom}{\mathsf{atom}}
\newcommand{\anelem}{\texttt{elem}}

\newcommand{\instrsetof}[1]{\mathsf{instrsOf}(#1)}

\newcommand{\coreof}[1]{\texttt{c}(#1)}
\newcommand{\physaddrof}[1]{\texttt{addr}(#1)}
\newcommand{\itypeof}[1]{\texttt{type}(#1)}
\newcommand{\pdatafrominitial}{\texttt{DataFromInitial}}

\newcommand{\readvalof}[1]{\texttt{loadval}(#1)}
\newcommand{\interpretationof}[1]{{#1}^{\mathcal{\anaxiomaticsem}}}
\newcommand{\nodeOf}[1]{\mathsf{node}(#1)}
\newcommand{\opof}[1]{\texttt{op}(#1)}
\newcommand{\labelof}[1]{\lambda(#1)}
\newcommand{\muhbgraph}{\mu\hbrel}
\newcommand{\anaxiomaticsem}{\mathcal{A}}

\newcommand{\psamecore}{\texttt{SameCore}}
\newcommand{\psameaddress}{\texttt{SameAddr}}
\newcommand{\piswrite}{\texttt{IsStore}}
\newcommand{\pisread}{\texttt{IsLoad}}

\newcommand{\ttrue}{\texttt{true}}

\newcommand{\transitionrelation}{\Delta}

\newcommand{\astate}{\mathsf{q}}

\newcommand{\stageS}{\text{\lstinline[style=axiom]{S}}}
\newcommand{\stageT}{\text{\lstinline[style=axiom]{T}}}
\newcommand{\aximposs}{\anaxiomaticsem^\#}

\newcommand{\setpredicates}{\mathbb{P}}
\newcommand{\apredicate}{\mathsf{P}}

\newcommand{\setactions}{\mathsf{Act}}

\newcommand{\rightact}{\mathsf{right}}
\newcommand{\stayact}{\mathsf{stay}}
\newcommand{\copyact}{\mathsf{sched}}
\newcommand{\yu}{\mathsf{U}}
\newcommand{\ve}{\mathsf{V}}
\newcommand{\qu}{\mathsf{q}}
\newcommand{\fst}[1]{\texttt{fst}(#1)}
\newcommand{\enabled}[1]{\mathsf{enabled}(#1)}
\newcommand{\aconf}{\gamma}
\newcommand{\aconfiguration}{\aconf}
\renewcommand{\init}{\mathsf{init}}

\newcommand{\anopmodel}{\mathcal{M}}
\newcommand{\tracesOf}[2]{\mathsf{traces}_{#1}(#2)}
\newcommand{\dropact}{\mathsf{drop}}

\newcommand{\diffr}{\mathsf{diff}_{\mathit{r}}}
\newcommand{\coupled}{\mathsf{coup}}

\newcommand{\predsof}[1]{\mathsf{non}\hbrel(#1)}

\newcommand{\setsymbinstr}{\mathtt{I}}
\newcommand{\setsymbevent}{\mathtt{E}}
\newcommand{\asymbevent}{\mathtt{e}}
\newcommand{\perm}{\mathsf{perm}}

\newcommand{\axexecstage}{\text{\lstinline[style=axiom]{Exe}}}
\newcommand{\axdecstage}{\text{\lstinline[style=axiom]{Dec}}}
\newcommand{\axfetstage}{\text{\lstinline[style=axiom]{Fet}}}
\newcommand{\axcomstage}{\text{\lstinline[style=axiom]{Com}}}

\newtheorem{assumption}{Assm.}

\newtheorem{claim}{Claim}
\newtheorem{theorem}{Theorem}
\newtheorem{lemma}{Lemma}
\newtheorem{definition}{Definition}
\newtheorem{example}{Example}
\newtheorem{corollary}{Corollary}

\SetAlgorithmName{Operational Loop}{heuristic}{List of Heuristics}

\lstdefinelanguage{smt}{
  sensitive = false,
  keywords={declare, fun, synth, check, assert, sat, blocking, define, constraint},
  numbers=left,
  numberstyle=\footnotesize,
  stepnumber=1,
  numbersep=8pt,
  showstringspaces=false,
  breaklines=true,
  frame=top,
  comment=[l]{;},
}

\lstdefinestyle{axiom}{
  belowcaptionskip=1\baselineskip,
  breaklines=true,
  xleftmargin=\parindent,
  showstringspaces=false,
  basicstyle=\bfseries\ttfamily,
  keywordstyle=\bfseries\color{green!40!black},
	keywordstyle = [2]{\sf\color{blue}},
  morekeywords={Cons, microops, exists, SameCore, SameAddr, microops, SameMicroop, IsAnyWrite, SameAddress},
  morekeywords=[2]{hb},
  commentstyle=\itshape\color{purple!40!black},
  identifierstyle=\color{violet},
  stringstyle=\color{orange},
}
\definecolor{mauve}{rgb}{0.58,0,0.82}

\lstdefinestyle{rvasm}{
  literate={ö}{{\"o}}1
           {ä}{{\"a}}1
           {ü}{{\"u}}1,
  basicstyle=\bfseries\footnotesize\ttfamily,                    
  breaklines=true,                              
  commentstyle=\itshape\color{green!50!black},  
  keywordstyle=[1]\color{blue!80!black},        
  keywordstyle=[2]\color{orange!80!black},      
  keywordstyle=[3]\color{red!50!black},         
  stringstyle=\color{mauve},                    
  identifierstyle=\color{teal},                 
  morekeywords=[1]{ 
    lb, lh, lw, lbu, lhu,
    sb, sh, sw,
    sll, slli, srl, srli, sra, srai,
    add, addi, sub, lui, auipc,
    xor, xori, or, ori, and, andi,
    slt, slti, sltu, sltiu,
    beq, bne, blt, bge, bltu, bgeu,
    j, jr, jal, jalr, ret,
    scall, break, nop
  },
  morekeywords=[2]{ 
    .align, .ascii, .asciiz, .byte, .data, .double, .extern,
    .float, .globl, .half, .kdata, .ktext, .set, .space, .text, .word
  },
  morekeywords=[3]{ 
    zero, ra, sp, gp, tp, s0, fp,
    t0, t1, t2, t3, t4, t5, t6,
    s1, s2, s3, s4, s5, s6, s7, s8, s9, s10, s11,
    a0, a1, a2, a3, a4, a5, a6, a7,
    ft0, ft1, ft2, ft3, ft4, ft5, ft6, ft7,
    fs0, fs1, fs2, fs3, fs4, fs5, fs6, fs7, fs8, fs9, fs10, fs11,
    fa0, fa1, fa2, fa3, fa4, fa5, fa6, fa7
  },
  tabsize=4,                                    
  showstringspaces=false                        
}

\definecolor{mygray}{gray}{0.6}

\makeatletter
\providecommand*{\ldash}{%
  \mathrel{%
    \mathpalette\@ldash\vdash
  }%
}
\newcommand*{\@ldash}[2]{%
  \reflectbox{$\m@th#1#2$}%
}
\makeatother

\newenvironment{cavresponse}
    {
      \begingroup
      \color{red}
    }
    {
      \endgroup
    }

\title{Automated Conversion of Axiomatic to Operational Models: Theory and Practice}

\author{
Adwait Godbole\IEEEauthorrefmark{1}, Yatin A. Manerkar\IEEEauthorrefmark{2}, and Sanjit A. Seshia\IEEEauthorrefmark{1} \\
\IEEEauthorblockA{
\IEEEauthorrefmark{1}\textit{University of California Berkeley, Berkeley, USA} $\quad$
\IEEEauthorrefmark{2}\textit{University of Michigan, Ann Arbor, USA}
}
}

\begin{document}
\pagenumbering{arabic}

\maketitle

\begin{abstract}
    A system may be modelled as an \emph{operational} model (which has explicit notions of state and transitions between states) or an \emph{axiomatic} model (which is specified entirely as a set of invariants). Most formal methods techniques (e.g., IC3, invariant synthesis, etc) are designed for operational models and are largely inaccessible to axiomatic models. Furthermore, no prior method exists to automatically convert axiomatic models to operational ones, so operational equivalents to axiomatic models had to be manually created and proven equivalent.
    
    In this paper, we advance the state-of-the-art in axiomatic to operational model conversion. We show that general axioms in the $\mu$spec axiomatic modelling framework cannot be translated to equivalent finite-state operational models. We also derive restrictions on the space of $\mu$spec axioms that enable the feasible generation of equivalent finite-state operational models for them. As for practical results, we develop a methodology for automatically translating $\mu$spec axioms to equivalent finite-state automata-based operational models. We demonstrate the efficacy of our method by using the models generated by our procedure to prove the correctness of ordering properties on three RTL designs.
\end{abstract}
\section{Introduction}

When modelling hardware or software systems using formal methods, 
one traditionally uses \emph{operational} models 
(e.g. Kripke structures~\cite{Baier2008PrinciplesOM}), 
which have explicit notions of state and transitions. 
However, one may also model a system \emph{axiomatically}, where 
instead of a state-transition relation,
the system is specified entirely by a set of axioms (i.e., invariants) that it maintains.
Executions that obey the axioms are allowed, and those that violate one or more axioms are forbidden.
The vast majority of formal methods work uses the operational modelling style. However, axiomatic models have been used to great effect in certain domains such as memory models,
where they have shown order-of-magnitude improvements in verification performance over equivalent operational models~\cite{herdingcats}.

Operational and axiomatic models each have their own advantages and disadvantages
\cite{manerkarThesis}. Operational models can be more intuitive 
as they typically resemble the system that they are modelling.
Hence one is not required to reason about invariants to write the model. 
On the other hand, axiomatic models tend to be more concise
and potentially offer faster verification~\cite{herdingcats}.

Many formal methods (e.g., refinement procedures~\cite{Burch1994AutomaticVO}, invariant synthesis, IC3/PDR~\cite{Bradley2011SATBasedMC,En2011EfficientIO}) are set up to use operational models. Axiomatic models are largely or completely incompatible with these techniques, 
as the axioms constrain full traces rather than a step of the transition relation.
One way to take advantage of these techniques when using axiomatic models is to create and use operational models equivalent to the axiomatic models. The only prior method of doing this was to first manually create the operational model and then manually prove it equivalent to the axiomatic model. There have been several works doing so \cite{herdingcats,Nienhuis2016AnOS,Lahav2016TamingRC,Owens2009ABX,Pulte2018SimplifyingAC}.

Manually creating an operational model and proving equivalence 
is cumbersome and error-prone. 
The ability to automatically generate operational models 
equivalent to a given axiomatic model 
would be beneficial,
eliminating both the time spent creating 
the operational model as well as the need for tedious manual equivalence proofs.
Generated models can then be fed into techniques currently requiring operational models (e.g. IC3/PDR).

To this end, we make advances in this paper towards the automatic conversion of axiomatic models to equivalent operational models, on both theoretical and practical fronts. In our work, 
we focus specifically on $\mu$spec~\cite{lustig:coatcheck}, a well-known axiomatic framework for modelling microarchitectural orderings,
which has been used in a wide range of contexts 
\cite{Lustig2014PipeCheckSA,Manerkar2017RTLCheckVT,Manerkar2018PipeProofAM,Trippel2018C,Manerkar2015CCICheckU} 
including memory consistency, cache coherence and hardware security.


On the theoretical front, we show that it is impossible to convert general $\mu$spec axioms to equivalent finite-state operational models.
However, we show that it is feasible to generate equivalent operational models for a specific subset of $\mu$spec (henceforth referred to as \uspecSub).
On the practical side, we develop a method to automatically translate 
axioms
in \uspecSub into equivalent finite-state operational models consisting of \emph{axiom automata} 
(finite automata that monitor whether an axiom has been violated).
Furthermore, for arbitrary $\mu$spec axioms, our method can generate 
operational models 
that are equivalent to the axioms
up to a program-size bound.

To evaluate our technique, we convert axioms 
for three RTL designs
to their corresponding operational models: 
an in-order multicore processor (\texttt{multi\_vscale}),
a memory-controller (\texttt{sdram\_ctrl}),
and an out-of-order single-core processor (\texttt{tomasulo}).
We showcase how the generated models can be used with procedures like BMC and IC3/PDR which are usually inaccessible for axiomatic models
and produce both bounded and unbounded proofs of correctness.

Overall, the contributions of this work are as follows:
\begin{itemize}
    \item We prove that generation of equivalent finite-state operational models for arbitrary $\mu$spec axioms is impossible.
    \item We provide a procedure for generating equivalent finite-state operational models for universal axioms in \uspecSub.
    \item We propose the \textit{axiom-automata} formulation to generate equivalent finite-state operational models from universal axioms in \uspecSub (or from arbitrary $\mu$spec axioms if only guaranteeing equivalence up to a bounded program size).
    \item We evaluate our method for operational model generation by using our generated models to prove the (bounded/unbounded) correctness of ordering properties on three RTL designs: \texttt{multi\_vscale}, \texttt{tomasulo}, and \texttt{sdram\_ctrl}.
\end{itemize}

\textbf{Outline.}
\S\ref{sec:uspec} covers the syntax and semantics of 
$\mu$spec used in this paper. 
\S\ref{sec:desires} covers the formulation 
of the space of operational models we consider. 
They have finite control-state and 
read-only input tapes for the instruction 
streams (programs) executed by each core.
\S\ref{sec:overview} defines our notions of soundness, completeness, and equivalence when comparing operational and axiomatic models. 
In \S\ref{sec:feasibility}, we show that it is infeasible to synthesise equivalent finite-state operational models from arbitrary axiomatic models. We develop an underapproximation, called $t$-reordering
boundedness, that addresses this by bounding the depth of reorderings possible.
In \S\ref{sec:extensibility} we restrict $\mu$spec further by requiring \emph{extensibility} (preventing current events from influencing orderings between previous events). 
Restricting $\mu$spec by 
$t$-reordering boundedness and extensibility is sufficient to enable the automatic generation of equivalent finite-state operational models (Thm. \ref{thm:tcompleteness}).
\S\ref{sec:operational} describes our conversion procedure based on axiom automata.
\S\ref{sec:experiments} evaluates our technique by using it to generate operational models, 
which are then used for checking correctness properties of RTL designs.
\S\ref{sec:related} covers related work, and \S\ref{sec:conclusion} concludes.
Additional material and proofs can be found in the supplementary material.




\begin{figure}
    \centering
    \resizebox{0.48\textwidth}{!}{
\begin{tikzpicture}[->, thick,
roundnode/.style={circle, draw=green!60, fill=green!5, very thick, minimum size=7mm},
squarednode/.style={rectangle, draw=red!60, fill=red!5, very thick, minimum size=5mm},
]
\def\pta{7}
\path[every node/.style={font=\sffamily\small}]
    (1.5,0.5) edge[bend left] node {} (2,1)
    (4.5,0.75) edge node {}  (4.5,0.5)
    (4.5,0) edge node {} (4.5,-0.25)
    (\pta,-0.5) edge[bend right] node {} (\pta+0.5,0)
    ;

\draw (0,0) rectangle node {\muspec} (1.5,0.5);
\draw (2,0.75) rectangle node {refinability (Def. \ref{def:refinable})} (\pta,1.25);
\draw (2,0) rectangle node{t-reordering bound (Def. \ref{def:treorder})} (\pta,0.5);
\draw (2,-0.75) rectangle node{extensibility (Def. \ref{def:extensibility})} (\pta,-0.25);
\draw (\pta+0.5,0) rectangle node {$\mu$specRE} (\pta+2,0.5);
\end{tikzpicture}
    }
    \caption{The relation between $\mu$spec and \uspecSub}
    \vspace{-0.6cm}
    \label{fig:roadmap}
\end{figure}

    
\textbf{Challenges.} 
    \label{para:challenges}
    While axiomatic models enforce constraints over 
    complete executions, operational models do this local
    to each transition. Ensuring that behaviours generated by 
    the latter is also allowed by the former
    requires performing non-local consistency checks 
    which are hard to reason about, especially for unbounded executions.
    This has been observed in manual operationalization works as well.
    Taking the example of
    \cite{Nienhuis2016AnOS}, (which operationalizes C11), we address issues of 
    eliminating consistent executions too early \cite[\S3]{Nienhuis2016AnOS}
    and repeatedly checking consistency \cite[\S4]{Nienhuis2016AnOS}
    by developing concepts such as t-reordering boundedness (Def. \ref{def:treorder}) and 
    extensibility (Def. \ref{def:extensibility}).
    Though we focus on \muspec{}, we believe many of the underlying challenges 
    and concepts carry over to frameworks 
    such as Cat \cite{herdingcats}.
\section{$\mu$spec Syntax and Semantics}
\label{sec:uspec}

\subsection{$\mu$spec Syntax}
\begin{figure}[h]
\begin{footnotesize}
\begin{align*}
\langle\anaxiom\rangle := ~~&\forall \asymbinstr ~~ \anaxiom \bnf \exists \asymbinstr ~~ \anaxiom \bnf \aformula(\asymbinstr_1, \cdots, \asymbinstr_m) \\
\langle\aformula\rangle := ~~&\aformula \land \aformula ~~|~~ \aformula \lor \aformula ~~|~~ \neg \aformula \bnf \langle\anatom\rangle \\
\langle\anatom\rangle := ~~&\asymbinstr_1 <_r \asymbinstr_2 \bnf \hbrel(\asymbinstr_1.\astage, \asymbinstr_2.\astage) \bnf \apredicate(\asymbinstr_1, \dots) \\ 
\langle \astage \rangle := ~~& \fetchstage ~~|~~ \decodestage ~~|~~ \execstage ~~|~~ \wbstage ~~|~~ \cdots
\end{align*}
\end{footnotesize}
\caption{$\mu$spec Syntax.}\
\label{fig:uhbgrammar}
\vspace{-0.4cm}
\end{figure}
$\mu$spec~\cite{lustig:coatcheck} is a domain-specific language used for specifying microarchitectural orderings.
A \muspec{} model consists of \textit{axioms} that 
enforce first-order constraints over execution graphs; 
each axiom quantifies over instructions and is required to 
be a sentence (not have any free variables).
Execution graphs that satisfy the axioms and 
are acyclic are deemed as valid executions.
%
While ISA-level models \cite{herdingcats,Manson2005TheJM,Sewell2010x86TSO}
treat single instructions as atomic entities, 
\muspec{} decomposes the execution of an instruction into a set of
atomic \textit{events}.
Each instruction $\aninstr$ and stage $\astage$ 
is associated with an event $\event{\aninstr}{\astage}$. 
A program execution is viewed as a directed acyclic graph
called a micro-architectural happens-before graph
($\muhbgraph$ graph) \cite{Lustig2014PipeCheckSA}.
Such a graph for a given program has nodes 
corresponding to events of form 
$\event{\aninstr}{\astage}$ for each instruction $\aninstr$ in the program and each stage $\astage$ prescribed by the model.
Edges in the graph correspond to the \textit{happens-before} ($\hbrel$)
relation: $\hbrel(\anevent_1, \anevent_2)$ says that $\anevent_1$ happened
before $\anevent_2$. Thus, a cyclic $\muhbgraph$ graph corresponds to an impossible scenario where an event happens before itself, and thus represents an execution that cannot occur on the microarchitecture. 

Fig.~\ref{fig:uhbgrammar} specifies $\mu$spec syntax. It has three types of atoms:
\begin{enumerate}[label=(\roman*)]
    \item $\hbrel(\asymbinstr_1.\astage, \asymbinstr_2.\astage)$: happens-before predicate
    \item $\aninstr_1 <_r \aninstr_2$: the reference order (typically the program order)
    \item $\apredicate(\asymbinstr_1, \dots)$: instruction predicate atoms
\end{enumerate}
The second atom captures order in which 
instructions appear in a given program thread.
The third are predicates over instructions which capture instruction properties, e.g. opcode, source/destination registers.

\begin{figure}
\begin{lstlisting}[style=axiom, mathescape=true, xleftmargin=0em]
ax0: fall i1. hb(i1.Exe,i1.Com) 
ax1: fall i1,i2. (i1$<_r$i2  $\land$ DepOn(i1,i2))       
    $\implies$ hb(i1.Exe,i2.Exe)
ax2: fall i1,i2. SameCore(i1,i2) $\Rightarrow$
(hb(i1.Exe,i2.Exe)$\lor$hb(i2.Exe,i1.Exe))
ax3: fall i1,i2. i1$<_r$i2 $\Rightarrow$ hb(i1.Com,i2.Com)
\end{lstlisting}
\caption{An example axiomatic model.}
\label{fig:ex-axmodel}
\end{figure}

\begin{figure*}[h]
    \centering 
    \vspace{-0.2cm}
    \begin{subfigure}{.45\textwidth}
        \centering
        \includegraphics[scale=0.2]{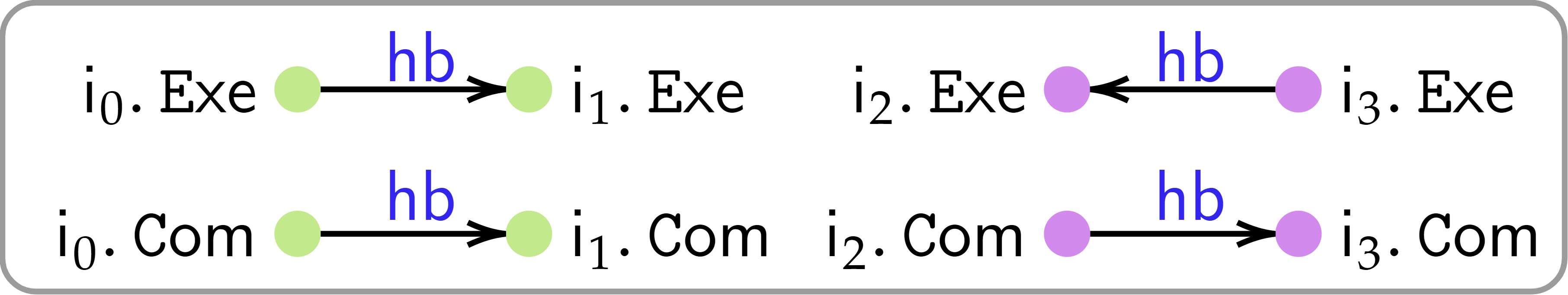}
        \vspace{-0.1cm}
        \caption{}
        \label{fig:runex2:a}
        \end{subfigure}
    \begin{subfigure}{.45\textwidth}
        \centering
        \includegraphics[scale=0.2]{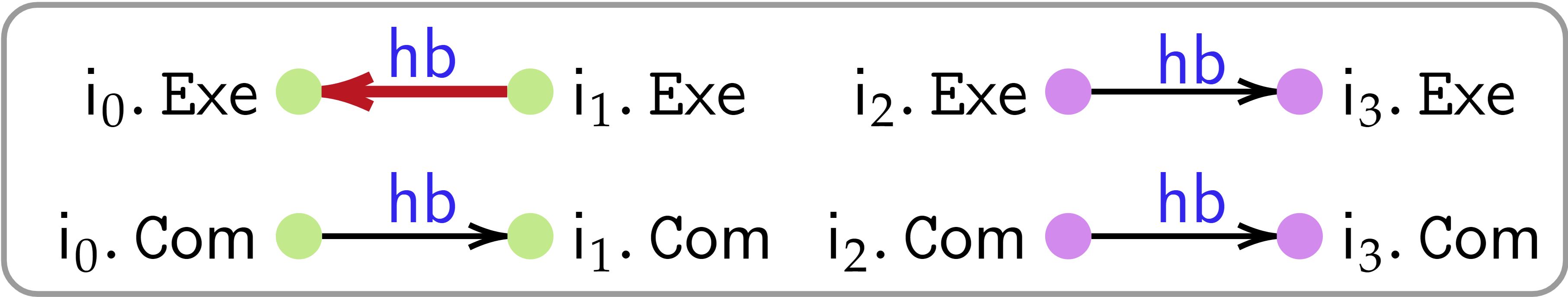}
        \vspace{-0.1cm}
        \caption{}
        \label{fig:runex2:b}
    \end{subfigure}
    \vspace{-0.1cm}
    \caption{Valid (a) and an invalid (b) execution graphs for program in Fig. \ref{fig:progsnip} and axioms in Fig \ref{fig:ex-axmodel}. The red edge violates \lstinline[style=axiom]{ax1}.}
    \label{fig:runex2}
\end{figure*}
\begin{figure}[h]
\centering
\vspace{-0.4cm}
\begin{tabular}{c || c}
\begin{lstlisting}[style=rvasm, mathescape]
$\aninstr_0$: lw  r1, 42(r0)  
$\aninstr_1$: add r3, r2, r1 
\end{lstlisting} & 
\begin{lstlisting}[style=rvasm, mathescape]
 $\aninstr_2$: lw  r4, 42(r0)
 $\aninstr_3$: add r3, r2, r1
\end{lstlisting}
\end{tabular}
\caption{Example program snippet}
\label{fig:progsnip} 
\vspace{-0.2cm}
\end{figure}

We identify two types of axioms of interest:
\textit{Universal axioms} are of the form: $\forall \asymbinstr_1 \cdots \forall \asymbinstr_k ~\phi(\asymbinstr_1, \cdots, \asymbinstr_k)$, 
and represent constraints applied symmetrically over all tuples of instructions in a program. 
\textit{Predicate-free axioms} are axioms that do not have occurences of predicate ($\apredicate$) atoms.
We extend these terms to an axiomatic semantics if all axioms are of that type.
In this work, our theoretical treatment focuses on universal semantics. 
Practically though, some underlying ideas
carry over to arbitrary axioms
as we discuss in \S\ref{sec:operational}, \S\ref{sec:experiments}.



\subsection{Illustrative \muspec{} Example}
\label{sec:uspec_example}

Consider the four axioms below. In the axioms, \lstinline[style=axiom]{i1, i2} are instruction variables and 
\lstinline[style=axiom]{Exe,Com} are stage names (short for execute and commit respectively).
The axiom \lstinline[style=axiom]{ax0} requires that for each instruction, the execute stage (\lstinline[style=axiom]{Exe}) of that instruction must happen before the commit stage (\lstinline[style=axiom]{Com}).
Intuitively, \lstinline[style=axiom]{ax1} says that when \lstinline[style=axiom]{i2}
depends on \lstinline[style=axiom]{i1}
(captured by the predicate \lstinline[style=axiom]{DepOn})
, \lstinline[style=axiom]{i1}
should be executed before \lstinline[style=axiom]{i2};
\lstinline[style=axiom]{ax2} says that the execute events
of instructions on the same core
should be \textit{totally ordered} by $\hbrel$.
The third axiom \lstinline[style=axiom]{ax3} says that when \lstinline[style=axiom]{i1}
and \lstinline[style=axiom]{i2} are in 
program order (denoted by $<_r$), \lstinline[style=axiom]{i1} must be committed
before \lstinline[style=axiom]{i2}.

Fig. \ref{fig:runex2} shows valid and invalid execution graphs for the
program snippet in Fig. \ref{fig:progsnip}.
The snippet is of a 2-core program, with two instructions per core.
Instruction 
$\aninstr_1$ is dependent on the result of
$\aninstr_0$ (since its source register is the same as the destination
of $\aninstr_0$).
In the example axiomatic semantics, \lstinline[style=axiom]{ax1}
requires that the execute event of instruction $\aninstr_0$ 
be before that of $\aninstr_1$.
The execution in Fig. \ref{fig:runex2:b} is invalid w.r.t. \lstinline[style=axiom]{ax1}
since $\event{\aninstr_1}{\axexecstage}$ is executed before $\event{\aninstr_0}{\axexecstage}$.
The execution in Fig. \ref{fig:runex2:a} is valid 
even though the $\event{\aninstr_2}{\text{\lstinline[style=axiom]{Exe}}}$ and 
$\event{\aninstr_3}{\text{\lstinline[style=axiom]{Exe}}}$ events are reordered
since $\aninstr_3$ does not depend on $\aninstr_2$.
Both executions are valid w.r.t. \lstinline[style=axiom]{ax2} and \lstinline[style=axiom]{ax3}.


\subsection{Programming Model}
We consider multi-core systems 
with each core executing a straightline program
over a finite domain of operations. 
This 
is common in
memory models \cite{herdingcats,Lustig2014PipeCheckSA,Manerkar2015CCICheckU,Batty2011MathematizingCC}
and
distributed systems \cite{Ahamad2005CausalMD} literature.

\label{subsec:programmingmodel}

\subsubsection{Cores}
The system consists of $n$ processor cores:
$\setcores = [n]$. Each core executes operations from a 
\textit{finite} set $\setoperations$. 
The axiomatic model $\anaxiomaticsem$ 
assigns predicates from $\setpredicates$
an interpretation over the universe $\setoperations$. 
We denote this interpretation as
$\apredicate^\anaxiomaticsem \subseteq \setoperations^k$
for an arity-$k$ predicate.

\subsubsection{Instruction streams} An \textit{instruction stream} $\istream$ is a 
word over $\setoperations$: $\istream \in \setoperations^*$. 
A program $\aprogram$ is a set of 
\textit{per-core} instruction streams: $\{\istream_\acore\}_{\acore\in \setcores}$.
For a core $\acore$ and label $0 \leq j < |\istream_\acore|$,
we call the triple $(\acore, j, \istream_\acore[j])$ an
\textit{instruction}\footnote{Note the terminology: operations are
commands that the core can execute.
Since we interpret predicates over $\setoperations$ we require $|\setoperations|$ to be a finite set for 
computability reasons.
Instructions are operations combined with 
the label and core identifier (and hence form an infinite set).}.
We denote components of instruction $\aninstr = (\acore, j, \istream_\acore[j])$, 
as: $\coreof{\aninstr} = \acore$, 
label $\labelof{\aninstr} = j$ and operation
$\opof{\aninstr} = \istream_\acore[j]$.
The set of instructions occuring in $\aprogram$ is:
$\instrsetof{\aprogram} = \{(\acore, j, \istream_\acore[j]) ~|~ \acore \in \setcores, 0 \leq j < |\istream_\acore|\}$
and the set of all possible instructions as 
$\setinstr = \setcores \times \mathbb{Z}^{\geq 0} \times \setoperations$.

\subsubsection{Instruction stages}
Instruction execution in \muspec{} is decomposed into stages.
The set of stages, $\setstages$, 
is a parameter of the semantics.
Instruction $\aninstr$ performing in stage $\astage$,
(i.e. $\event{\aninstr}{\astage}$)
is an atomic event in an execution. 
The execution of $\aprogram$ 
is composed of the set of events:
$\eventsOf{\aprogram} = \{\aninstr.\astage ~|~ \aninstr \in \instrsetof{\aprogram}, \astage \in \setstages\}$.
The set of all possible events is $\setevents = \{\aninstr.\astage ~|~ \aninstr \in \setinstr, \astage \in\setstages\}$.

\begin{definition}[Event]
   An event $\anevent$ is of the form $\event{\aninstr}{\astage}$. 
   It represents the instruction $\aninstr \in \setinstr$, (atomically) performing in stage $\astage\in\setstages$.
\end{definition}


\begin{example}
    Following the example in Fig. \ref{fig:progsnip} 
    we consider an architecture with two opcodes: 
    \lstinline[style=rvasm]{add, lw} for add and load 
    respectively. For each of these, we may have several
    actual operations (with different operands), thus giving us
    the set $\setoperations$. The program, $\aprogram$, in Fig. 
    \ref{fig:progsnip} has two cores: $\setcores = \{\acore_0, \acore_1\}$ and four instructions:
    $\instrsetof{\aprogram} = \aninstr_{\{0,1,2,3\}}$. We have, for example,
    $\coreof{\aninstr_1} = \acore_0, \labelof{\aninstr_1} = 1$ while $\coreof{\aninstr_2} = \acore_1, \labelof{\aninstr_2} = 0$.
    
    Now we can consider a 4-stage
    microarchitecture with $\setstages = \{\emph{\axfetstage}, \emph{\axdecstage}, \emph{\axexecstage}, \emph{\axcomstage}\}$. The events for program $\aprogram$ 
    are $\eventsOf{\aprogram} = \{\event{\aninstr_0}{\emph{\axfetstage}}, \event{\aninstr_0}{\emph{\axdecstage}}, \cdots, \event{\aninstr_3}{\axcomstage} \}$ with 
    $|\eventsOf{\aprogram}| = 4\times 4 = 16$.
\end{example}

\subsection{Formal $\mu$spec Semantics}
\label{sec:uspec_semantics}

We now define the formal semantics $\mu$spec axioms. 

\begin{definition}[$\muhbgraph$ graph]
    For a program $\aprogram$, a $\muhbgraph$ graph
    is a directed acyclic graph, $G(V,E)$, with nodes $V = \eventsOf{\aprogram}$ representing events and edges representing the happens-before relationships, 
    i.e. $(\anevent_1, \anevent_2) \in E \equiv \hbrel(\anevent_1, \anevent_2)$.
\end{definition}





\noindent\textbf{Validity of $\muhbgraph$ graph w.r.t. an axiomatic semantics:}
Consider an axiomatic semantics $\anaxiomaticsem$ 
(i.e. a set of axioms).
A $\muhbgraph$ graph $G = (V, E)$
is said to represent a valid execution of program $\aprogram$ under $\anaxiomaticsem$ if
it satisfies all the axioms in $\anaxiomaticsem$.
We denote the validity of a $\muhbgraph$ graph $G$ by $G \models_\aprogram \anaxiomaticsem$.

\subsubsection*{Satisfaction w.r.t. an axiom} 


We first define satisfaction for the quantifier-free part, starting at the atoms. Let $s: \symbinstrset(\anaxiom) \rightarrow \setinstr$
be an instruction assignment for the symbolic instruction variables in axiom $\anaxiom$.

\smallskip
$
\begin{array}{l}
G\models \asymbinstr_1[s] <_r \asymbinstr_2[s] \iff \coreof{s(\asymbinstr_1)} = \coreof{s(\asymbinstr_2)} \\
\qquad\quad  ~\land~ \labelof{s(\asymbinstr_1)} < \labelof{s(\asymbinstr_2)} \qquad ... (i)  \\[0.1cm]
G\models \apredicate(\asymbinstr_1, \cdots, \asymbinstr_m)[s] \iff \\
\qquad\quad (\opof{s(\asymbinstr_1)}, \cdots, \opof{s(\asymbinstr_m)}) \in \interpretationof{\apredicate} \qquad ... (ii) \\[0.1cm]
G\models \hbrel(\asymbinstr_1.\astage_1, \asymbinstr_2.\astage_2)[s] \iff \\
\qquad\quad(\event{s(\asymbinstr_1)}{\astage_1},
\event{s(\asymbinstr_2)}{\astage_2}) \in E^+ \qquad ...(iii)
\end{array}
$
\smallskip

In \textit{(i)} The reference order $<_r$ relates instructions $\aninstr_1, \aninstr_2$ 
from the same instruction stream if $\aninstr_1$ is before $\aninstr_2$.
In \textit{(ii)} we extend predicate interpretations, $\apredicate^\anaxiomaticsem$, (defined over $\setoperations$)
to instructions by taking the $\opof{\cdot}$ component.
Finally, $\hbrel$ atoms are interpreted as $E^+$, i.e. transitive closure of $E$, as stated in \textit{(iii)}.
Operators $\land, \lor, \neg$ have their usual semantics.

\smallskip
$
\begin{array}{l  l  l}
G \models_\aprogram \phi[s] &\equiv G \models \phi[s] \\
&\quad \text{ for quantifier-free } \phi \\
G \models_\aprogram \forall \asymbinstr~ \aformula[s] &\equiv G \models_\aprogram \aformula[s[\asymbinstr\leftarrow \aninstr]] 
\\
&\quad \text{ for all } \aninstr \in \instrsetof{\aprogram} \setminus \rangeOf{s} \\
G \models_\aprogram \exists \asymbinstr~ \aformula[s] &\equiv G \models_\aprogram \aformula[s[\asymbinstr \leftarrow \aninstr]] 
\\
&\quad \text{ for some } \aninstr \in \instrsetof{\aprogram} \setminus \rangeOf{s} 
\end{array}
$
\smallskip

We define the satisfaction of a (quantified) 
axiom $\anaxiom$ by a graph $G$,
denoted by $G \models_\aprogram \anaxiom$ above.
The base case is $G \models_\aprogram \phi[s]$
(where $\phi$ is quantifier-free) and follows the 
earlier definitions. 
We extend $G \models_\aprogram \phi$ with
(almost) usual quantification semantics:
$\forall$ ($\exists$) quantifies over all (some) instructions in $\instrsetof{\aprogram}$,
except that we only consider distinct variable assignments\footnote{
This is largely a syntactic convenience for our technical treatment, and does not change 
the expressivity in any way.}.
Execution $G$ is a valid execution of $\aprogram$ under semantics $\anaxiomaticsem$, 
$G \models_\aprogram \anaxiomaticsem$, 
if $G \models_\aprogram \anaxiom$ for all axioms in $\anaxiomaticsem$.
\newcommand{\transsys}{\mathcal{TS}}
\newcommand{\setstates}{\mathcal{Q}}

\section{Operational Model of computation}
\label{sec:desires}




To concretize our claims,
we introduce a model of computation that
characterizes the space of models of interest.
We choose to focus on finite-state operational models that 
generate totally ordered traces, where transitions 
represent single ($\event{\aninstr}{\astage}$) events.
While there are less restrictive models
(e.g. event structures \cite{Moiseenko2020ReconcilingES,Jeffrey2016OnTA}), 
such models require
specialized, typically under-approximate, verification techniques 
(e.g. \cite{Norris2013CDScheckerCC,Aronis2018EffectiveTF,Kokologiannakis2020HMCMC}).
Our choice is motivated by the ability to 
(a) have finite-state implementations of generated models (e.g. in RTL) 
and 
(b) verify against these models with off-the-shelf tools.




\subsection{Model of computation}

Intuitively, the model of computation resembles a 1-way transducer 
\cite{Berstel1979TransductionsAC,Sakarovitch2009ElementsOA} with 
multiple (read-only) input tapes (corresponding to instruction streams).
This allows us to execute programs of unbounded length with 
a finite control.\footnote{A Kripke 
structure-based formalism is insufficient since we want to execute unbounded programs 
with distinguished instructions without explicitly modelling control logic.}

\subsubsection{Model definition}
An operational model is parameterized by cores $\setcores$,
stages $\setstages$, and a history parameter $h \in \nat \cup \{\infty\}$
which bounds the length of tape to the left of the head.
It is a tuple $(\setstates, \transitionrelation, \astate_\init, \astate_\final)$:
\begin{itemize}
    \item $\setstates$ is a finite set of control states
    \item $\transitionrelation \subseteq \setstates \times (\setinstr\cup\{\ldash\})^{|\setcores|} \times \setstates \times \setactions$ is the transition relation where $\setactions$ is the set of actions
    \item $\astate_\init \in \setstates$ is the initial state
    \item $\astate_\final \in \setstates$ is the final state which must be absorbing
\end{itemize}
A model is finite-state if 
$\setstates$ is finite and that it has bounded-history if $h\in\nat$.
For the end goal of effective verification, 
we are interested in finite-state, bounded-history models
since it is precisely such 
models that can be compiled to finite-state systems.

\subsubsection{Model semantics}
A configuration is a triple $\gamma = (\yu, \qu, \ve)$
where $\yu : \setcores \rightarrow \setinstr^*$, 
$\ve : \setcores \rightarrow \setinstr^*$
and $\qu \in \setstates$. 
Intuitively $\yu$ ($\ve$) represent, for each instruction stream, 
the contents of the input tape to the left (right) of the head respectively.
For a bounded history machine, a configuration is \textit{allowed} only if 
$|\yu(\acore)| \leq h$ for all $\acore \in \setcores$ (for unbounded history all configurations are allowed).

The set of actions is 
\begin{align*}
    \setactions = ~& 
\{\rightact(\acore) ~|~ \acore \in \setcores\} ~\cup \\
& \{\stayact\} ~\cup \\
& \{\copyact(\acore, i, \astage) ~|~ \acore \in \setcores, \astage \in \setstages, i \in [h]\} ~\cup \\
& \{\dropact(\acore, i) ~|~ \acore \in \setcores, i \in [h]\}
\end{align*}
Intuitively, these represent in order: 
motion of the tape head for $\acore$ to the right,
silent (no-effect),
generation of an event,
removing the $i^{th}$ instruction from the left of the head.
We provide full semantics in the supplementary material.

For word $w \in \setinstr^*$, $\fst{w}$ is its first element
if $w \neq \epsilon$ and $\ldash$ otherwise.
Transitions are conditioned on the 
instructions that the tape-heads point to: transition
$(\astate_1, (\aninstr_1, \cdots, \aninstr_{|\setcores|}), \astate_2, \_) \in \Delta$ is enabled in configuration
$\aconfiguration = (\yu, \astate, \ve)$
if 
$\astate_1 = \astate$ and $\fst{\ve(\acore)} = \aninstr_\acore$ for each $\acore \in \setcores$.



\subsubsection{Runs}
The initial configuration is given by $\aconfiguration_\init(\aprogram) = (\yu_\init, \astate_\init, \ve_\init)$ 
where $\yu_\init = \lambda \acore.~\epsilon$ and $\ve_\init = \lambda \acore.~\istream_\acore$, i.e. for each core, the left of the tape head is empty, and the right of the tape head consists of the instruction stream for that core.  
Starting from $\aconfiguration_\init(\aprogram)$, the machine transitions
according to the transition rules.
Such a sequence of configurations
$\aconf_\init(\aprogram) = \aconf_0 \xrightarrow{\anevent_1} \aconf_1 \cdots \xrightarrow{\anevent_m}
\aconf_m$, where all $\aconf_i$ are allowed is called a run. 
A run is called accepting if it ends in the state $\astate_\final$.

\subsubsection{Traces}
\label{subsubsec:traces}
The sequence of event labels $\atrace = \anevent_1\cdots\anevent_m$ 
annotating a run is the \textit{trace} corresponding to the run. 
Each label is an event from $\setevents$
and hence $\sigma \in \setevents^*$.
We view $\atrace$ as a (linear) $\muhbgraph$ execution graph
$\anevent_1 \hbarrow \anevent_2 \cdots \hbarrow \anevent_m$, 
and hence define $\atrace \models \anaxiomaticsem$ in the usual way.
Accordingly, we will sometimes refer to $\atrace$ as an execution of a program $\aprogram$.
The set of traces
corresponding to accepting runs 
of an operational model $\anopmodel$ on a program $\aprogram$
are denoted as $\tracesOf{\anopmodel}{\aprogram} \subseteq \setevents^*$.


\section{Soundness, Completeness, and Equivalence}
\label{sec:overview}

We proceed to formalize the notion of equivalence
that relates axiomatic and operational models.
In literature \cite{litmus,herdingcats}, 
ISA-level behaviours of programs have been annotated by the 
read values of $\aload$ operations. Hence, one notion of equivalence 
might be to require that identical read values be possible
between 
the models.
While this may be reasonable for ISA-level behaviours, 
it can hide micro-architectural features:
different micro-architectural executions can have
identical architectural results. 
Given that \muspec{} models model executions at the 
granularity of microarchitectural events, we adopt a stronger notion of equivalence.
For soundness, we require that the operational semantics generates linearizations of $\muhbgraph$ graphs
that are valid under the axiomatic semantics. Formally:

\begin{definition}[Soundness]
\label{req:sound}
An operational model $\anopmodel$ is sound,
w.r.t. $\anaxiomaticsem$
if for any program $\aprogram$, each trace in $\tracesOf{\anopmodel}{\aprogram}$
is a linearization of some $\muhbgraph$ graph that is
valid under $\anaxiomaticsem$.
\end{definition}

Before defining completeness, we need to address a subtlety.
Since operational executions are 
viewed as $\muhbgraph$ graphs by interpreting trace-ordering
as the $\hbrel$ ordering, 
the operational model always generates linearized $\muhbgraph$ graphs.
However, in general, linearizations of valid $\muhbgraph$ graphs 
could end up being invalid w.r.t the axioms.
Consider Example \ref{ex:nonref}.
\begin{example}[Non-refinable axiom]
\normalfont
For the following axiom with $\setstages = \{\stageS\}$ 
the graph (a) is a valid execution.
However both of its linearizations (b) and (c) are invalid.
Thus, \textit{all} of the (totally-ordered) traces generated by our operational models will be deemed invalid under the axiomatic semantics.
This renders a direct comparison between operational
and axiomatic executions infeasible.
\begin{lstlisting}[style=axiom,mathescape]
fall i1,i2.($\neg$hb(i1.S,i2.S)$\land\neg$hb(i2.S,i1.S))
\end{lstlisting}
\begin{figure}[h]
    \centering
    \vspace{-0.4cm}
    \includegraphics[scale=0.5]{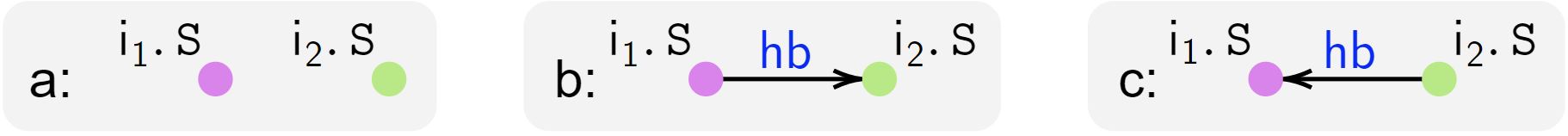}
    \vspace{-0.3cm}
\end{figure}
\label{ex:nonref}
\end{example}

To address this issue, we develop the notion of refinability.
For two $\muhbgraph$ graphs $G = (V, E)$ and $G' = (V', E')$, we say that
$G'$ refines $G$, denoted
$G \sqsubseteq G'$ if (1) $V = V'$ and 
(2) $(\anevent_1, \anevent_2) \in E^+ \implies (\anevent_1, \anevent_2) \in E'^+$.
\begin{definition}[Refinable $\hbrel$]
    \label{def:refinable}
	An axiomatic semantics $\anaxiomaticsem$ is refinable
    if for any program $\aprogram$, and $\muhbgraph$ graph $G$ s.t. $G \models_\aprogram \anaxiomaticsem$, 
    we have $G' \models_\aprogram \anaxiomaticsem$ for \emph{all} linear graphs $G'$ satisfying $G \sqsubseteq G'$.
\end{definition}
Refinability says that all
linearizations of a valid graph are valid.
While executions under axiomatic semantics are 
given by (partially-ordered) $\muhbgraph$ graphs, 
our class of operational models generate totally-ordered traces.
Refinability bridges this gap by relating valid 
$\muhbgraph$ graphs to valid traces. 
Interestingly, we can check whether a universal axiomatic semantics 
satifies refinability, which at a high level, 
we show via a small model property (Lemma \ref{lem:checkref}).
Due to space constraints, we defer the proof to the supplement.
\begin{lemma}
    Given a universal axiomatic semantics we can
    determine whether the semantics is refinable.
    \label{lem:checkref}
\end{lemma}

Refinability is especially important for completeness.
For non-refinable semantics, 
validity of linearizations 
cannot be checked based on the
axioms, as all linearizations may be invalid (Example \ref{ex:nonref}). 
\begin{quote}
\textit{We assume that the semantics satisfies refinability.}
\end{quote}
We define completeness and our formal problem statement.
\begin{definition}[Completeness]
\label{req:complete}
An operational model $\anopmodel$ is complete, if
for any program $\aprogram$ and valid $\muhbgraph$ graph $G \models_\aprogram \anaxiomaticsem$, 
$\tracesOf{\anopmodel}{\aprogram}$ contains all linearizations of $G$.
\end{definition}

\noindent\textbf{Formal Problem Statement}
Given an axiomatic semantics, $\anaxiomaticsem$, a set of cores $\setcores$ 
and stages $\setstages$, generate a \emph{finite state, bounded history} model, 
$\anopmodel = (\setstates, \transitionrelation, \astate_\init)$, 
which satisfies soundness and completeness (Defns. \ref{req:sound} and \ref{req:complete}).
\section{Enabling Synthesis by Bounding Reorderings}
\label{sec:feasibility}

In this section, we develop some theoretical results for the 
synthesis of operational models. First, we show that 
synthesis of sound and complete (viz. Defn. \ref{req:sound} and \ref{req:complete}) 
finite-state operational models is not possible.
Then we provide an underapproximation for the completeness requirement,
called $t$-completeness,
that enables the synthesis of finite-state models. 
This still does not allow for bounded-history models as 
future events can influence past orderings (Example \ref{ex:nfe}).
In \S\ref{sec:extensibility} we add \textit{extensibility} thus enabling both finite-state and bounded-history models,
our original goal.


\subsection{An impossibility result}
We show that it is in fact impossible to develop a finite-state transition system $\anopmodel$ that 
satisfies the requirements prescribed in Defns. \ref{req:sound} and \ref{req:complete}.
Figure~\ref{fig:nofinite} gives an axiomatic semantics $\aximposs$ 
(with $\setstages = \{\stageS, \stageT\}$)
such that for all possible finite-state models, there is some program such that either 
soundness or completeness is violated. 
\begin{figure}[h]
\centering
\begin{lstlisting}[style=axiom, mathescape]
ax0: fall i1.    hb(i1.S,i1.T)
ax1: fall i1,i2. hb(i1.S,i2.S)$\Rightarrow$hb(i1.T,i2.T)
\end{lstlisting}
\caption{Semantics $\aximposs$ that does not allow bounded synthesis}
\label{fig:nofinite}
\end{figure}
In words, the axioms in Fig. \ref{fig:nofinite} state the following constraints: 
\lstinline[style=axiom]{ax0} says that for each instruction,
the \lstinline[style=axiom]{S} stage event happens before the \lstinline[style=axiom]{T} stage, 
and
\lstinline[style=axiom]{ax1} enforces that for any two instructions, the ordering between their \stageS{}
stage events implies an identical ordering between their \lstinline[style=axiom]{T} stage events. We have the following:
\begin{theorem}
	\label{thm:infeasible}
	For a single-core program $\aprogram$ with an instruction stream of 
	$|\istream_{\acore_1}| = m$ instructions, there is no model
	$\anopmodel = (\setstates, \transitionrelation, \astate_\init, \astate_\final)$ that is sound and complete 
	w.r.t. $\aximposs$ and $\aprogram$, and s.t. $|\setstates| < \mathcal{O}(2^m/m)$, even with $h = \infty$.
\end{theorem}
We provide an intuitive explanation, deferring details to the supplement.
In valid executions of $\aximposs$, 
\stageS{} stage events can be ordered arbitrarily, 
while \stageT{} stage events must maintain
the same ordering as that of corresponding \stageS{} stages. 
Hence the machine must remember the \stageS{} orderings in its finite control.
However, the number of such orderings grows 
(exponentially) with the number of instructions $m$, 
implying that existence of a finite-state model that works for all programs is not possible.
\begin{corollary}
    \label{corr:imposs}
	There does not exist a finite state operational model (even with $h = \infty$)
	which is sound and complete with respect to the $\anaxiomaticsem^\#$ axioms.
\end{corollary}


\newcommand{\idxstart}{\mathsf{start}}
\newcommand{\idxend}{\mathsf{end}}
\newcommand{\idxpfxend}{\mathsf{pfxend}}

\subsection{An underapproximation result}
Given the results of the previous section, we must relax 
some constraint imposed on the operationalization:
we choose to relax completeness.
To do so, we define an under-approximation called \textit{$t$-reordering bounded traces}.
Intuitively, this imposes two constraints:
(a) it bounds the depth of reorderings between instructions on each core,
(b) it bounds the skew between instructions across two cores.

For two instructions $\aninstr_1, \aninstr_2$ on the same core, 
let $\diffr(\aninstr_1, \aninstr_2) = \labelof{\aninstr_2} - \labelof{\aninstr_1}$ (recall that $\labelof{\aninstr}$ is the instruction index of $\aninstr$).
Consider a trace $\atrace$ of program $\aprogram$. For $\aninstr \in \instrsetof{\aprogram}$, 
we define the starting index of $\aninstr$, denoted as $\idxstart(\aninstr)$,
as the smallest $1 \leq j \leq |\atrace|$ such that $\anevent_j = \event{\aninstr}{\astage}$
($\anevent_j$ was the first event for instruction $\aninstr$ in $\atrace$).
Similarly we define the ending index, $\idxend(\aninstr)$ as the largest index for some event of $\aninstr$.
Let the \textit{prefix-closed end index} of $\aninstr$ be the max of $\idxend$ over 
instructions that are $\leq_r \aninstr$:
$\idxpfxend(\aninstr) = \max\{\idxend(\aninstr') ~|~ \aninstr' \leq_r \aninstr\}$.
Two instructions $\aninstr_1$ and $\aninstr_2$ are coupled in a trace 
(denoted as $\coupled(\aninstr_1, \aninstr_2)$) if the intervals
$[\idxstart(\aninstr_1), \idxpfxend(\aninstr_1)], [\idxstart(\aninstr_2), \idxpfxend(\aninstr_2)]$ overlap.
\begin{definition}[$t$-reordering bounded traces]
	\label{def:treorder}
    A trace is $t$-reordering bounded if, for any pair of instructions $\aninstr_1, \aninstr_2$ with
	$\coreof{\aninstr_1} = \coreof{\aninstr_2}$, 
	(1) if $\event{\aninstr_2}{\astage_2} \hbarrow \event{\aninstr_1}{\astage_1}$
		then $\diffr(\aninstr_1, \aninstr_2) < t$ and
	(2) if $\coupled(\aninstr_1, \aninstr), \coupled(\aninstr, \aninstr_2)$
		for some $\aninstr$
		then $|\diffr(\aninstr_1, \aninstr_2)| < t$.
\end{definition}
Intuitively, (1) says that an instruction cannot be reordered with another 
that precedes it by $\geq t$ indices, while
(2) says that instructions
on a core cannot be \textit{stalled} while more than 
$t$ instructions are executed on another.
Note that t-reordering boundedness is a property of traces, 
and not of axioms.
We now relax completeness (and hence equivalence)
to require that the operational model at least generate all
$t$-reordering bounded linearizations
(instead of all linearizations).

\newtheorem{innercustomreq}{Definition}
\newenvironment{customreq}[1]
  {\renewcommand\theinnercustomreq{#1}\innercustomreq}
  {\endinnercustomreq}

\begin{customreq}{5*}[$t$-completeness]\label{req:treorder}
    An operational model $\anopmodel$ is $t$-complete w.r.t. 
	an axiomatic model $\anaxiomaticsem$, if for each program $\aprogram$ and 
	$G\models_\aprogram \anaxiomaticsem$, $\tracesOf{\anopmodel}{\aprogram}$ contains
	all $t$-reordering bounded linearizations of $G$.
\end{customreq}

Replacing Defn. \ref{req:complete} 
with its $t$-bounded relaxation (Defn. \ref{req:treorder})
addresses the issue 
of having to keep track of an unbounded number of orderings.
However, to allow for finite implementations in practice,
in addition to finite-state, we also require bounded-history ($h \in \nat$).
This is addressed in the next section.

\section{Adding Extensibility}
\label{sec:extensibility}

As illustrated by the following example, 
the $t$-reordering bounded underapproximation 
is insufficient to achieve 
bounded-history operational model synthesis on its own.

\begin{example}[Need for extensibility]
\normalfont
	Consider a single stage axiomatic semantics: $\setstages = \{\stageS\}$, and predicate $\setpredicates = \{\apredicate\}$.
\begin{lstlisting}[style=axiom,mathescape]
fall i0,i1,i2. ($\apredicate$(i0,i1,i2) $\land$ i0$<_r$i1) 
                $\implies$ $\neg$hb(i1.S,i0.S)
\end{lstlisting}
	There cannot be a sound, $t$-complete, 
	and bounded-history (for bound $h$)
	model for this axiom (for some $t > 1$). 
	To see this, consider a (single-core) program $\aprogram$,
	with instructions $\aninstr_0 \cdot \aninstr_1 \cdots \aninstr_{h+1}$.
	Depending on the instructions in $\aprogram$, 
	the interpretation $\apredicate^\anaxiomaticsem$ 
	of $\apredicate$ can either be
	(a) $\apredicate^\anaxiomaticsem = \{(\aninstr_0, \aninstr_1, \aninstr_{h+1})\}$ or
	(b) $\apredicate^\anaxiomaticsem = \phi$.
	In the former case, the ordering 
	$\event{\aninstr_1}{\stageS} \hbarrow \event{\aninstr_0}{\stageS}$ is invalid while in the latter it is valid.
	Since we only allow a $h$-sized history, 
	$\event{\aninstr_0}{\stageS}$ must be scheduled
	before the tape-head reaches $\aninstr_{h+1}$, 
	i.e. before the machine can determine
	which of (a)/(b) hold. 
	Since the machine cannot determine whether events 
	$\event{\aninstr_0}{\stageS}$, $\event{\aninstr_1}{\stageS}$ 
	can be reordered, this leads either to a model which is 
	unsound (always reorders) or incomplete (never reorders).
	\label{ex:nfe}
\end{example}



\newcommand{\ispref}[1]{{\color{blue!50} #1}}
\newcommand{\ispost}[1]{{\color{red!50} #1}}

\begin{figure}
	\centering 
	\includegraphics[scale=0.5]{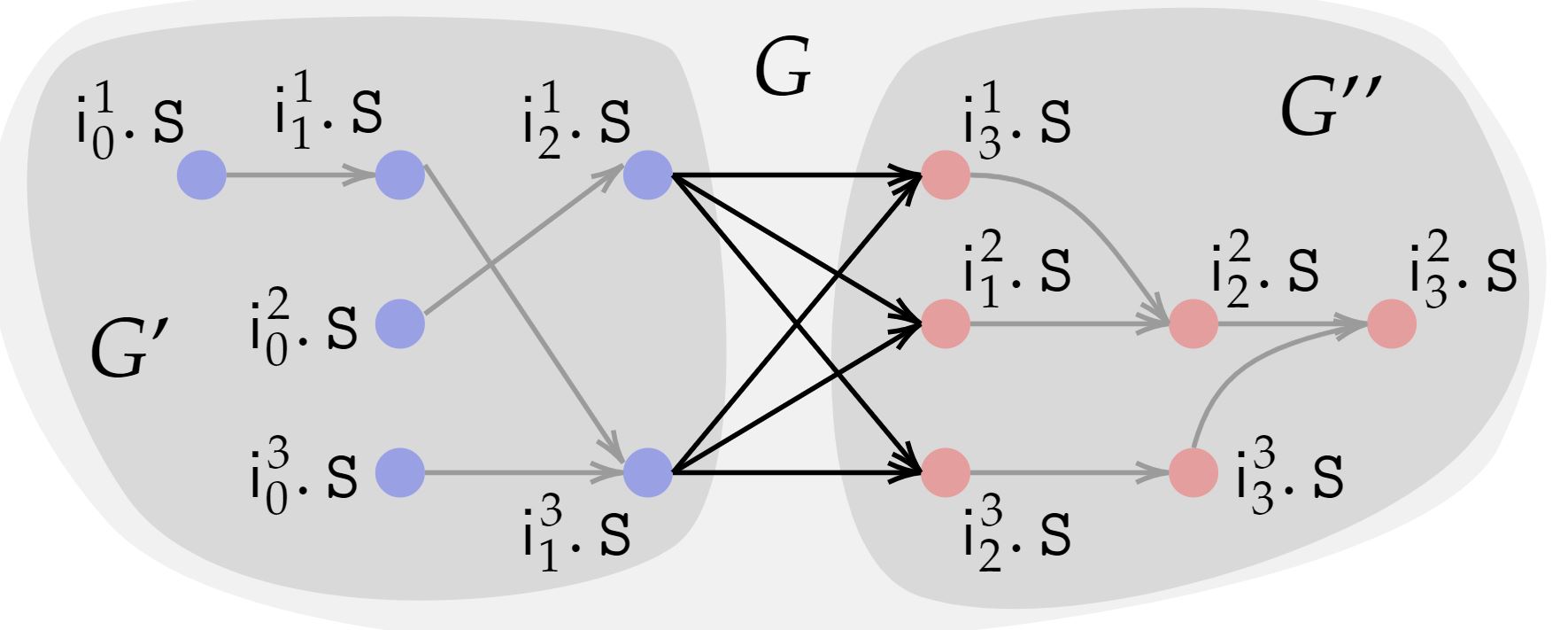}
\caption{
$\aprogram$ has instruction streams $\ispref{\aninstr^1_0}\cdot\ispref{\aninstr^1_1}\cdot\ispref{\aninstr^1_2}\cdot\ispost{\aninstr^1_3}$,
$\ispref{\aninstr^2_0}\cdot\ispost{\aninstr^2_1}\cdot\ispost{\aninstr^2_2}\cdot\ispost{\aninstr^2_3}$, and 
$\ispref{\aninstr^3_0}\cdot\ispref{\aninstr^3_1}\cdot\ispost{\aninstr^3_2}\cdot\ispost{\aninstr^3_3}$.
Blue instructions form the prefix ${\color{blue!50}\aprogram'}$ (i.e. ${\color{blue!50}\aprogram'} \preceq \aprogram$) and red
its residual ${\color{red!50}\aprogram''} = \aprogram \oslash {\color{blue!50}\aprogram'}$.
The figure shows executions $G'$ of ${\color{blue!50}\aprogram'}$, 
$G''$ of ${\color{red!50}\aprogram''}$, and their composition $G = G' \triangleright G''$.
}
\label{fig:extensibility}
\end{figure}


Thus, we need an additional restriction to enable generation of 
operational models with a finite history parameter $h$. 
We propose extensibility, which 
intuitively states that partial executions 
of program $\aprogram$
that have not violated any axioms 
can be composed with valid executions 
of the residual program, 
to generate valid complete executions of $\aprogram$.
To do this, we extend the notion of validity to partial executions through \textit{prefix programs}.

A program $\aprogram$ can be split into a 
prefix $\aprogram'$ (blue) and the residual suffix $\aprogram''$ (red)
(Fig. \ref{fig:extensibility}).
Formally, $\aprogram'$ is a \textit{prefix} of program $\aprogram$,
if $\aprogram'$ has instruction streams $\{\istream'_i\}$,
each of which is a prefix of the instr. streams $\{\istream_i\}$ of $\aprogram$.
We denote that $\aprogram'$ is a prefix of $\aprogram$ by $\aprogram' \preceq \aprogram$. 
For programs $\aprogram$, $\aprogram'$ such that $\aprogram' \preceq \aprogram$ we denote 
the \textit{residual} of $\aprogram$ w.r.t. $\aprogram'$ as $\aprogram'' = \aprogram \oslash \aprogram'$.
$\aprogram'$ has instr. streams $\istream_\acore''$:
for each core $\acore$, $\istream_\acore = \istream_\acore' \cdot \istream_\acore''$.

In Fig. \ref{fig:extensibility}, for example,
the first instruction stream of $\aprogram$ is
$\ispref{\aninstr^1_0}\cdot\ispref{\aninstr^1_1}\cdot\ispref{\aninstr^1_2}\cdot\ispost{\aninstr^1_3}$.
The prefix program ${\color{blue!50}\aprogram'}$ has (the prefix) $\ispref{\aninstr^1_0}\cdot\ispref{\aninstr^1_1}\cdot\ispref{\aninstr^1_2}$ as its first instr. stream. 
On the other hand, the residual program, 
${\color{red!50}\aprogram''} = \aprogram\oslash{\color{blue!50}\aprogram'}$, has the suffix $\ispost{\aninstr^1_3}$
as its instruction stream.

For graphs $G' = (V', E')$ and $G'' = (V'', E'')$, with $V' \cap V'' = \emptyset$ 
we define $G' \triangleright G''$ as the graph $G = (V, E)$ where, 
(1) $V = V' \cup V''$, and
(2) $E = E' \cup E'' \cup \{(\anevent', \anevent'') ~|~ \anevent' \in \sinkv(E'), \anevent'' \in \sourcev(E'')\}$.
The example in Fig. \ref{fig:extensibility}
illustrates such a composition:
we have $G = G' \triangleright G''$.


\begin{definition}[Extensibility]
	\label{def:extensibility}
    An axiom $\anaxiom$ satisfies extensibility if 
    for any programs $\aprogram$ and $\aprogram'$ s.t. $\aprogram' \preceq \aprogram$, 
    and $\aprogram'' = \aprogram \oslash \aprogram'$ if $G' \models_{\aprogram'} \anaxiom$ 
    and $G'' \models_{\aprogram''} \anaxiom$ 
    then $G' \triangleright G'' \models_\aprogram \anaxiom$.
	An axiomatic semantics $\anaxiomaticsem$ satisfies extensibility if all axioms $\anaxiom \in \anaxiomaticsem$
	satisfy extensibility.
\end{definition}
We require that the axiomatic model satisfies extensibility.
We define \uspecSub (RE stands for Refinable, Extensible) as the subset of $\mu$spec
in which all axioms are universal, refinable, and extensible.
Finite-state, bounded-history synthesis is feasible for axioms 
in \uspecSub, as we discuss in the next section.
Like refinability, we can check whether an axiom can satisfy
extensibility (Lemma \ref{lem:checkext}).
We provide a proof in the supplement.
\begin{lemma}
	\label{lem:checkext}
    Given a universal axiom we can determine whether is satisfies extensibility.
\end{lemma}

\newcommand{\anautomaton}{\mathcal{A}}
\newcommand{\atransitionrel}[1]{\xrightarrow{#1}}
\newcommand{\stateset}{\mathcal{Q}}
\newcommand{\acc}{\mathsf{acc}}
\newcommand{\acceptstate}{q^{\acc}}
\newcommand{\rejectstate}{q^{\rej}}

\newcommand{\windowmap}{\mathcal{W}}
\newcommand{\window}{\mathsf{w}}
\newcommand{\windowof}[1]{\windowmap_{#1}}

\newcommand{\apredatom}{\alpha}

\newcommand{\activeauto}{\mathsf{active}}
\newcommand{\kcoup}{\mathsf{AC}_k}
\newcommand{\activeinst}[1]{\mathsf{AC}_#1}
\newcommand{\cmall}{\mathsf{CM}}
\newcommand{\nfall}{\mathsf{NF}}
\newcommand{\cmtpoint}{\mathsf{cp}}
\newcommand{\fetpoint}{\mathsf{fp}}
\newcommand{\ipset}{\mathsf{IP}}
\newcommand{\cmset}{\mathsf{pCM}}
\newcommand{\nfset}{\mathsf{pNF}}

\section{Converting to Operational Models Using Axiom Automata}
\label{sec:operational}

In this section, we describe our approach
that converts an axiomatic model $\anaxiomaticsem$ in \uspecSub
into a sound and $t$-complete, finite-state, bounded-history 
operational model $\anopmodel$.
We focus on a single universal axiom $\forall \asymbinstr_1, \cdots, \asymbinstr_k \phi$,
but this can be easily extended to a set of axioms.
For our axiomatic conversion, we develop axiom automata - 
automata that check for axiom compliance as the operational model executes.

\subsection{Axiom Automata}
\label{subsec:aauto}
In what follows, we fix a (universal) axiom 
$\anaxiom = \forall \asymbinstr_1 \cdots \forall \asymbinstr_k ~ \phi(\asymbinstr_1, \cdots, \asymbinstr_k)$,
and let $\setsymbinstr(\anaxiom) = \{\asymbinstr_1, \cdots, \asymbinstr_k\}$,
$\setsymbevent(\anaxiom) = \{\event{\asymbinstr}{\astage} ~|~ \asymbinstr \in \setsymbinstr(\anaxiom), \astage \in \setstages\}$.
Let $\predsof{\anaxiom}$ denote the non-$\hbrel$ atoms in $\phi$, i.e. instruction predicate applications 
and $<_r$ orderings. 
A \textit{context} is a map $\acontext: \predsof\anaxiom \rightarrow \mathbb{B}$ that assigns a 
true/false value to each element of $\predsof\anaxiom$.
A context $\acontext$
\textit{agrees with} an assignment $s : \setsymbinstr(\anaxiom) \rightarrow \setinstr$ if 
the evaluation of each non-$\hbrel$ atom under $s$ matches $\acontext$.
Let $\acontext(s)$ be the context agreeing with $s$.
We extend $s$ to events: for $\asymbevent = \event{\asymbinstr}{\astage}$, $s(\asymbevent) = \event{s(\asymbinstr)}{\astage}$ 
and words over events: for $w \in \setsymbevent(\anaxiom)^*$, 
$s(w) = s(w[0])\cdots s(w[|w|-1]) \in \setevents^*$.
As mentioned in \S\ref{subsubsec:traces}, we interpret $w \in \setevents^*$ as the $\muhbgraph$ graph 
$w[0] \hbarrow w[1] \cdots \hbarrow w[|w|-1]$.
For a finite alphabet $S$, we denote the set of permutations over $S$ as $S^\perm$.

Lemma \ref{lem:aauto} says that given $\anaxiom$ and a context $\acontext$, 
we can construct a finite state automaton
that recognizes acceptable orderings of $\setsymbevent(\anaxiom)$. 
This follows since the set $\setsymbevent(\anaxiom)$ (and hence
the desired language) is finite.
The main observation behind Lemma \ref{lem:aauto} is that once the interpretation 
of the $\predsof{\anaxiom}$ atoms is fixed, 
the allowed orderings 
can be represented as a language 
over the symbolic events $\setsymbevent(\anaxiom)$.

\begin{lemma}[Axiom-Automata]
    \label{lem:aauto}
    Given axiom $\anaxiom$ and context $\acontext$, 
    there exists a finite-state automaton $\axiomautoof{\anaxiom[\acontext]}$ 
    over alphabet $\setsymbevent(\anaxiom)$
    with language 
    $\{w ~|~ w\in \setsymbevent(\anaxiom)^\perm, s(w) \models \phi(\asymbinstr_1, \cdots, \asymbinstr_k)[s]
    ~~\text{ for all } s \text{ that agree with }\acontext\}$.
\end{lemma}


\newcommand{\aautoset}{\mathcal{C}}

\subsubsection{Concretization of an axiom automaton}
The automaton $\axiomautoof{\anaxiom[\acontext]}$
recognizes words over symbolic events $\setsymbevent(\anaxiom)$.
Given an assignment $s$, 
we denote the (\textit{concretized}) automaton for $s$ w.r.t $\anaxiom$ as $\axiomautoof{\anaxiom, s}$.
This automaton is isomorphic to $\axiomautoof{\anaxiom[\acontext(s)]}$,
with the alphabet $\setsymbevent(\anaxiom)$ replaced by its image $s(\setsymbevent(\anaxiom))$ under $s$. 
For $I \subseteq \setinstr$,
we denote by $\axiomautoof{\anaxiom, I}$ as the set of axiom automata over $I$:
$\{\axiomautoof{\anaxiom, s} ~|~ s: \setsymbinstr(\anaxiom) \rightarrow I\}$.

\subsubsection{A basic operationalization}
\label{subsubsec:basicop}
Lemma \ref{lem:aauto} suggests an operationalization for $\anaxiom$.
For a program $\aprogram$, if a trace 
$\atrace$ is accepted by all (concrete) automata 
$\axiomautoof{\anaxiom, \instrsetof{\aprogram}}$ 
then $\atrace \models \phi[s]$ holds for each  assignment $s$,
thus satisfying $\anaxiom$.
%
Since the number of these automata ($|\axiomautoof{\anaxiom, \instrsetof{\aprogram}}|
\sim |\instrsetof{\aprogram}|^k$ for an axiom with $k$ quantified variables)
increases with $\aprogram$, the model is not finite state.
Even so, this enables us to construct operational models 
\textit{for a given bound on $|\instrsetof{\aprogram}|$},
even for non-universal axioms
(by converting existential quantifiers into finite
disjunctions over $\instrsetof{\aprogram}$).
We demonstrate an application of this in \S\ref{sec:experiments}, 
where we check that a processor satisfies an axiom ensuring 
correctness of read values.

\subsection{Bounding the number of active instructions}
Generating all concrete automata (statically) does not give us
a finite state model. 
We need to bound the number of automata
maintained at any 
point in the trace.

\begin{figure}[h]
    \centering
    \includegraphics[scale=0.23]{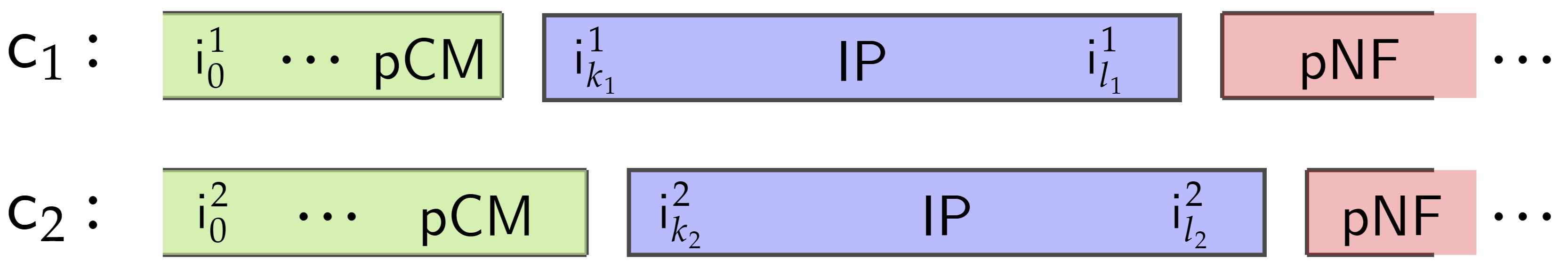}
    \caption{Completed prefix ($\cmset$), in-progess ($\ipset$) and not-fetched postfix ($\nfset$) of
    instructions during execution. 
    }
    \label{fig:pipeline}
    \vspace{-0.3cm}
\end{figure}
For a $t$-reordering bounded trace
$\atrace$
of a program $\aprogram$ and
a trace index $0 \leq j \leq |\atrace|$,
let $\cmall(j)$ and $\nfall(j)$ 
be instructions which 
have executed all and none of their events
at $\atrace[j]$ respectively. 
We define the following auxillary terms:
\[
\begin{array}{rl}
    \cmset(j) &= \{\aninstr ~|~ \forall \aninstr'.~ \aninstr' \leq_r \aninstr \implies \aninstr' \in \cmall(j)\} \\
    \nfset(j) &= \{\aninstr ~|~ \forall \aninstr'.~ \aninstr \leq_r \aninstr' \implies \aninstr' \in \nfall(j)\} \\
    \ipset(j) &= \instrsetof{\aprogram} \setminus (\cmset(j) \cup \nfset(j))
\end{array}
\]
Intuitively $\cmset(j)$ represents the prefix-closed set of \textit{completed} instructions, 
$\nfset(j)$ represents the postfix-closed set of \textit{unfetched} instructions,
and $\ipset(j)$ are the rest (see Fig. \ref{fig:pipeline}).
By the first condition of $t$-reordering boundedness,
in-progress ($\ipset$) instructions
on each core are bounded by $t$ for all $j$:
\begin{lemma}
    \label{lem:ipbounded}
    For any $t$-reordering bounded trace $\atrace$, for all $0 \leq j \leq |\atrace|$, 
    we have, $|\ipset(j)| \leq |\setcores|\cdot t$.
\end{lemma}

\paragraph*{$\kcoup(j)$ instructions}
Two instructions $\aninstr, \aninstr'$ 
are $k$-\textit{coupled} in $\atrace$
if 
$\coupled(\aninstr, \aninstr_1), \coupled(\aninstr_1, \aninstr_2), \cdots, 
\coupled(\aninstr_{k-1}, \aninstr')$
for some $\aninstr_1, \aninstr_2, \cdots \aninstr_{k-1}$.
For trace $\atrace$ and $0 \leq j \leq |\atrace|$, we define
$k$-\textit{active} instructions at $j$, $\kcoup(j)$
as instructions from $\cmset(j) \cup \ipset(j)$ which are $k$-coupled with some instruction from $\ipset(j)$.

\begin{lemma}
    \label{lem:activeinst}
    For each $k$, there is a (program-independent) bound $b_k$, s.t. for any
    $t$-reordering bounded trace $\atrace$, for all $0 \leq j \leq |\atrace|$,
    we have $|\kcoup(j)| \leq b_k$.
\end{lemma}

Intuitively, an active instruction is one for which we still need to preserve 
ordering information to ensure the soundness of the operational model.
Lemmas \ref{lem:ipbounded}, \ref{lem:activeinst} imply that at all points in the trace, 
it suffices to maintain 
(1) a bounded history (containing $\ipset$ instructions) and 
(2) finite (function of $b_k$) state capturing information about all 
active instructions relevant to execution at that point. Finally, we get the main result.

\begin{theorem}
    \label{thm:tcompleteness}
    For a (refinable) universal axiomatic semantics that satisfies extensibility, 
    synthesis of finite-state, bounded-history operational models satisfying Def. 
    \ref{req:sound} and \ref{req:treorder} is feasible.
\end{theorem}
\section{Case Studies}
\label{sec:experiments}
\newcommand{\vscale}{\texttt{vscale}}
\newcommand{\multivscale}{\texttt{multi\_vscale}}
\newcommand{\tomasuloproc}{\texttt{tomasulo}}
\newcommand{\sdram}{\texttt{sdram\_ctrl}}

In this section, we demonstrate applications of operationalization. 
We discuss three case studies:
(1) $\multivscale$ is a multi-core extension of the 3-stage in-order
$\vscale$ \cite{vscale} 
processor,
(2) $\tomasuloproc$ is an out-of-order 
processor based on \cite{tomasulo}, and
(3) $\sdram$ is an SDRAM-controller \cite{sdram}.

For each case, we instrument the hardware designs by exposing ports that signal the
execution of events (e.g. \texttt{PC} ports in Fig. \ref{fig:expt}).
We convert axioms into an operational model $\anopmodel$ 
based on the approach discussed in \S\ref{sec:operational}. $\anopmodel$
is compiled to RTL and is synchronously composed with the hardware design,
where it transitions on the exposed event signals.
Thus, any violating behaviour of the hardware will lead $\anopmodel$ into a non-accepting 
($\texttt{bad}$) state.
Hence by specifying $\texttt{bad}$ as a safety property, we can perform verification
of the RTL design w.r.t. the axioms.
The operationalization approach 
enables us to perform both bounded and unbounded verification
using off-the-shelf hardware model checkers.
We highlight that this would not have been possible without operationalization.

We use the Yosys-based \cite{Wolf2013YosysAFV} 
SymbiYosys as the model-checker, with boolector \cite{boolector} and abc \cite{abcTool} as backend solvers for BMC and PDR proof strategies respectively.
Experiments are performed on an Intel Core i7 machine with
16GB of RAM.
We use our algorithm to automatically generate axiom automata,
but the compilation of the generated automata to RTL and 
their instrumentation with the design is done manually.
The experimental designs are available at \href{https://github.com/adwait/axiomatic-operational-examples}{https://github.com/adwait/axiomatic-operational-examples}.

\textbf{Highlights.}
We demonstrate how the operationalization framework enables us
to leverage off-the-shelf model checking tools implementing 
bounded and (especially) unbounded proof techniques such as IC3/PDR.
This would not have been possible directly with axiomatic models.
Even when Thm. \ref{thm:tcompleteness} does not apply
(e.g. non-universal/non-extensible axioms), following \S\ref{subsubsec:basicop}
we can fall back on a BMC-based check over all possible programs
under a bound on $|\instrsetof{\aprogram}|$.

\subsection{The $\multivscale$ processor}

\paragraph{Pipeline axioms on a single core}

We begin with the single-core variant of $\multivscale$. 
We are interested in verifying the pipeline axioms for this core. 
The first axiom states that pipeline stages must be in 
\lstinline[style=axiom]{Fet-DX-WB} order and the second enforces 
in-order fetch.

\begin{lstlisting}[style=axiom, mathescape]
ax1: fall i1. (hb(i1.Fet,i1.DX) $\land$ 
        hb(i1.DX,i1.WB))
ax2: fall i1,i2.i1$<_r$i2 $\Rightarrow$ hb(i1.Fet,i2.Fet)
\end{lstlisting}

\begin{figure}
    \centering
    \includegraphics[scale=0.4]{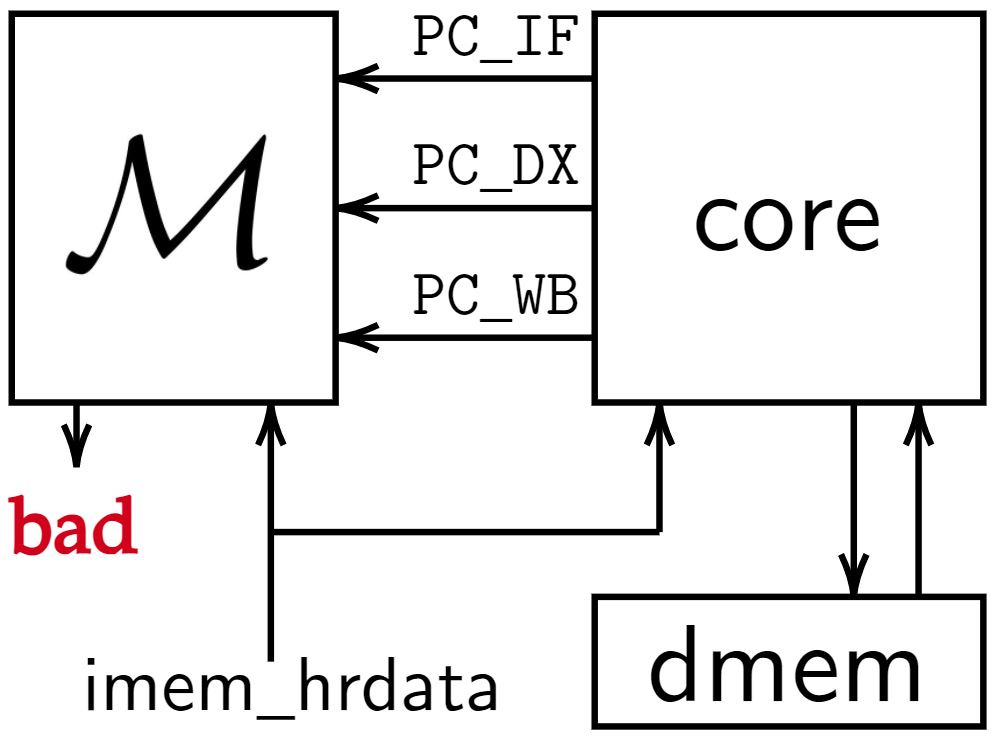}
    \caption{Experimental setup.}
    \label{fig:expt}
\end{figure}

The setup schematic is in Figure \ref{fig:expt}: $\anopmodel$
is the operational model implemented in RTL 
(note that we could do this only because the model is finite state 
and requires a finite history $h$). 
Given that it is a 3-stage in-order processor, 
at any given point each core has at most 3 instructions
in its pipeline and we can safely choose a history parameter of $h = 3$,
and $\anopmodel$ is complete for a reordering bound of $t = 3$.
We replace the \texttt{imem\_hrdata} (instruction data) 
connection to the core by an input signal that we can symbolically constrain.
Using this input signal, we can control the program 
(instruction stream) executed by the core.

Verification is performed with a PDR based proof
using the \texttt{abc pdr} backend. 
We experiment with various choices of 
instructions fed to the processor (by symbolically constraining $\texttt{imem\_hrdata}$).
In Fig. \ref{fig:exptres} below, we show the constraint and its PDR proof runtime, with
BMC runtime (depth = 20) for comparison. These examples demonstrate our ability 
to prove unbounded correctness.

\begin{figure}
    \centering
    \vspace{-0.2cm}
    \begin{tabular}{c  c  c}
        Instructions & PDR & BMC ($d=20$)  \\ \hline\hline
        ALU-R & 1m46s & 14m30s \\[3pt]
        ALU-I & 2m11s & 11m31s \\[3pt]
        Load+Store & 2m18s & 13m35s \\[3pt]
    \end{tabular}
\vspace{-0.15cm}
\caption{Proof runtimes for (\text{\lstinline[style=axiom, mathescape]{ax1}} $\land$ \text{\lstinline[style=axiom, mathescape]{ax2}}).
}
\vspace{-0.2cm}
\label{fig:exptres}
\end{figure}

\paragraph{Memory ordering on multi-core}

We now configure the design with 2 cores: $\acore_0$, $\acore_1$,
both initialized with symbolic load and store operations. 
We then perform verification w.r.t. the 
\lstinline[style=axiom]{ReadValues} (\lstinline[style=axiom]{RV}) axiom
shown below. This axiom says that for any read instruction 
(\lstinline[style=axiom]{i1}), the value read should be
the same as the most recent write instruction (\lstinline[style=axiom]{i2})
on the same address, or it should be the initial value.

\begin{lstlisting}[style=axiom, mathescape]
RV: fall i1,exists i2,fall i3. IsRead(i1) $\implies$ 
  (DataInit(i1)$\lor$(IsWrite(i2)$\land$
    SameAddr(i1,i2)$\land$hb(i2.DX,i1.DX)
    $\land$ValEq(i1,i2)$\land$((IsWrite(i3)$\land$SameAddr(i1,i3))$\implies$
  (hb(i3.DX,i2.DX)$\lor$hb(i1.DX,i3.DX)))))
\end{lstlisting}
This not a universal axiom, and hence Thm. \ref{thm:tcompleteness}
does not apply.
However, for bounded programs we can construct
$|\instrsetof\aprogram|^2$ concrete automata as discussed in \S\ref{subsubsec:basicop}. 
We convert the existential quantifier over \lstinline[style=axiom]{i2}
into a finite disjunction over $\instrsetof{\aprogram}$.
We perform BMC queries for programs with 
$|I| = |\instrsetof{\aprogram}| =  4, 6, 8$.

\begin{figure}[h]
    \centering
    \begin{tabular}{c  c  c c}
        $|I|$ & $|AA|$ & BMC $d$ & Time \\ \hline\hline
        4 & 16 & 12 & 3m10s    \\[3pt]
        6 & 36 & 16 & 15m48s    \\[3pt]
        8 & 64 & 20 & 1h58m
    \end{tabular}
\caption{Proofs runtimes for the Read-Values
axiom for different instruction counts ($|I|$).
}
\end{figure}
By keeping instructions symbolic, we effectively prove
correctness for \textit{all} programs 
within our bound $|I|$.
The table alongside shows the instruction bound, $|I|$, 
the number of axiom automata $|AA|$, 
BMC depth $d$, and runtime.
Though our theoretical results apply to universal axioms, 
this shows how an axiom automata-based operationalization can be applied to arbitrary axioms by
bounding $|\instrsetof{\aprogram}|$. 

\subsection{An OoO processor: $\tomasuloproc$}

Our second design is an out-of-order processor
(based on \cite{tomasulo})
that implements Tomasulo's algorithm.
The processor has stages: 
\lstinline[style=axiom]{F} (fetch), 
\lstinline[style=axiom]{D} (dispatch), 
\lstinline[style=axiom]{I} (issue), 
\lstinline[style=axiom]{E} (execute), 
\lstinline[style=axiom]{WB} (writeback), 
and
\lstinline[style=axiom]{C} (commit).
We verify in-order-commit, 
program-order fetch, and pipeline order axioms for this processor.
A BMC proof (with $d = 20$) takes $\sim$2m.

The axiom \lstinline[style=axiom]{axDep} given below 
is crucial for correct execution in an OoO processor. 
It enforces that \lstinline[style=axiom]{E} stages
for consecutive instructions
should be in program order
if the destination of the first instruction is
same as the source of the second,
i.e. dependent instructions are executed in order.
\begin{lstlisting}[style=axiom, mathescape]
axDep: fall i1, i2, (i1$<_r$i2 $\land$ Cons(i1,i2) $\land$ 
    DepOn(i1,i2)) $\implies$ hb(i1.E,i2.E)
\end{lstlisting}
We add 
a program counter ($\texttt{pc}$) to instructions
and define
$\text{\lstinline[style=axiom]{Cons}}(\aninstr_1, \aninstr_2) 
\equiv \texttt{pc}(\aninstr_1) + 4 = \texttt{pc}(\aninstr_2)$
and
$\text{\lstinline[style=axiom]{DepOn}}(\aninstr_1, \aninstr_2) \equiv  
\destof{\aninstr_1} = \srconeof{\aninstr_2} \lor \destof{\aninstr_1} = \srctwoof{\aninstr_2}$.

As before, 
we compose the
operational model $\anopmodel$ corresponding to this axiom 
with the RTL design.
We symbolically constrain the processor to 
execute a sequence of symbolic (\texttt{add} and \texttt{sub}) instructions and assert \texttt{!bad}.
A BMC query ($d = 20$) 
results in an assertion violation.
We manually identified the bug as being caused by the incorrect 
reset of entries in the Register Alias Table
in the $\axcomstage$ stage.
When committing instruction $\aninstr_0$, the entry
RAT$(\destof{\aninstr_0})$ is reset, while some instruction 
$\aninstr_1$ with $\destof{\aninstr_0} = \destof{\aninstr_1}$ is issued at the same cycle.
A third instruction $\aninstr_2$ with $\srconeof{\aninstr_2} = \destof{\aninstr_0}$ 
then reads the result of $\aninstr_0$ instead of $\aninstr_1$,
violating the axiom. We fix this bug and perform a BMC proof ($d = 20$), 
which takes $\sim$6m30s. This demonstrates how our techinique 
can be used to identify a bug, correct it and check the fixed design.

\subsection{A memory controller: $\sdram$}

To demonstrate the versatility of our approach,
we experiment with an SDRAM controller
\cite{sdram},
which interfaces a processor host with an SDRAM device,
with a ready-valid interface for read/write requests.
All intricacies related to interfacing with the SDRAM
are handled by maintaining appropriate control state
in the controller.
In the following, we once again convert 
axioms into an operational model
by our technique, 
and compose the generated model with the design.

First we verify pipeline-stage axioms for $\sdram$ for
write (4-stages) and read (5-stages) 
operations executed by the host interface.
A PDR-based (unbounded) proof
for the pipeline axioms requires $\sim$8m.
SDRAM requires a periodic refresh operation \cite{Jacob2007MemorySC}. 
The controller ensures that the host-level
behaviour is not affected by refreshes
by creating an illusion of atomicity for writes and reads.
This results in the axiom that 
once a write or read operation is underway,
no refresh stage should execute before it is completed.
We once again prove this property with PDR
which takes $\sim$1m30s.


\section{Related work}
\label{sec:related}


There has been much work on developing 
axiomatic (declarative) models for memory consistency in parallel systems, 
at the ISA level~\cite{herdingcats,Shasha1988EfficientAC,RVManual}, 
the microarchitectural level~\cite{Lustig2014PipeCheckSA,Manerkar2015CCICheckU,Lustig2016COATCheckVM}, 
and the programming language level~\cite{Batty2011MathematizingCC,Batty2015OverhaulingSA,Vafeiadis2015CommonCO,Watt2020RepairingAM,Lahav2017RepairingSC}. 
There has also been work on constructing equivalent operationalizations for these models, 
e.g., for Power~\cite{herdingcats}, ARMv8~\cite{Pulte2018SimplifyingAC}, RA\cite{Lahav2016TamingRC}, C++~\cite{Nienhuis2016AnOS}, and TSO \cite{Sewell2010x86TSO,Owens2009ABX}.
These constructions are accompanied by hand-written/theorem-prover based proofs,
demonstrating equivalence with the axiomatic model.
In principle, our work is related to these, however we enable \emph{automatic} 
generation of equivalent operational models from axiomatic ones, eliminating most of the manual effort.
At an abstract level, we have been inspired by classical works that have developed
connections between logics and automata \cite{Bchi1990OnAD,WolperVardiSistla83}.
In terms of the application to proving properties, the work closest to ours is 
RTLCheck \cite{Manerkar2017RTLCheckVT}, which compiles constraints from \muspec{}
to SystemVerilog assertions. 
These assertions are checked on a per-program basis. 
On the other hand, we demonstrate the ability to 
prove unbounded correctness. Additionally, for axioms that 
are not generally operationalizable (for unbounded programs), 
we demonstrate the ability to generate an operational model for some apriori known 
bound on the program size. In this case, we can verify correctness for 
\textit{all} programs of size upto that bound as opposed to on a per-program basis as RTLCheck do.
RTL2$\mu$spec \cite{Hsiao2021SynthesizingFM} aims to perform the reverse conversion: from RTL to 
\muspec{} axioms.

\section{Conclusion}
\label{sec:conclusion}

%
In this paper we make strides towards enabling greater interoperability between operational and axiomatic models, both through theoretical results and case studies.
We derive \uspecSub, a restricted subset of the $\mu$spec domain-specific language for axiomatic modelling.
We show that the generation of an 
equivalent finite-state operational 
model is impossible for general $\mu$spec axioms, 
though it is feasible for universal axioms in \uspecSub.
From a practical standpoint, we develop an approach based on axiom automata that enables us to automatically generate such equivalent operational models for universally quantified axioms in \uspecSub{} (or for arbitrary $\mu$spec axioms if equivalence up to a bound is sufficient).

The challenges we surmount for our conversion 
(discussed in \S\ref{para:challenges}) find parallels in 
manual operationalization works \cite{Nienhuis2016AnOS} and we believe that
the above concepts can be extended to formalisms such as Cat \cite{herdingcats}.


Our practical evaluation illustrates 
the key impact of this work---its ability to enable users of axiomatic models to take advantage of the vast number of techniques that have been developed for operational models in the fields of formal verification and synthesis.

\bibliographystyle{unsrt}
\bibliography{refs}

\begin{thebibliography}{10}

\bibitem{Baier2008PrinciplesOM}
Christel Baier and Joost-Pieter Katoen.
\newblock Principles of model checking.
\newblock 2008.

\bibitem{herdingcats}
Jade Alglave, Luc Maranget, and Michael Tautschnig.
\newblock Herding cats: Modelling, simulation, testing, and data mining for
  weak memory.
\newblock {\em ACM Trans. Program. Lang. Syst.}, 36(2), jul 2014.

\bibitem{manerkarThesis}
Yatin~A. Manerkar.
\newblock {\em Progressive Automated Formal Verification of Memory Consistency
  in Parallel Processors}.
\newblock PhD thesis, Princeton University, Princeton, NJ, USA, 2020.

\bibitem{Burch1994AutomaticVO}
Jerry~R. Burch and David~L. Dill.
\newblock Automatic verification of pipelined microprocessor control.
\newblock In {\em CAV}, 1994.

\bibitem{Bradley2011SATBasedMC}
Aaron~R. Bradley.
\newblock Sat-based model checking without unrolling.
\newblock In {\em VMCAI}, 2011.

\bibitem{En2011EfficientIO}
Niklas E{\'e}n, Alan Mishchenko, and Robert~K. Brayton.
\newblock Efficient implementation of property directed reachability.
\newblock {\em 2011 Formal Methods in Computer-Aided Design (FMCAD)}, pages
  125--134, 2011.

\bibitem{Nienhuis2016AnOS}
Kyndylan Nienhuis, Kayvan Memarian, and Peter Sewell.
\newblock An operational semantics for c/c++11 concurrency.
\newblock In {\em OOPSLA}, 2016.

\bibitem{Lahav2016TamingRC}
Ori Lahav, Nick Giannarakis, and Viktor Vafeiadis.
\newblock Taming release-acquire consistency.
\newblock {\em Proceedings of the 43rd Annual ACM SIGPLAN-SIGACT Symposium on
  Principles of Programming Languages}, 2016.

\bibitem{Owens2009ABX}
Scott Owens, Susmit Sarkar, and Peter Sewell.
\newblock A better x86 memory model: x86-tso.
\newblock In {\em TPHOLs}, 2009.

\bibitem{Pulte2018SimplifyingAC}
Christopher Pulte, Shaked Flur, Will Deacon, Jon French, Susmit Sarkar, and
  Peter Sewell.
\newblock Simplifying arm concurrency: multicopy-atomic axiomatic and
  operational models for armv8.
\newblock {\em Proceedings of the ACM on Programming Languages}, 2:1 -- 29,
  2018.

\bibitem{lustig:coatcheck}
Daniel Lustig, Geet Sethi, Margaret Martonosi, and Abhishek Bhattacharjee.
\newblock {COATCheck: Verifying Memory Ordering at the Hardware-OS Interface}.
\newblock In {\em 21st International Conference on Architectural Support for
  Programming Languages and Operating Systems (ASPLOS)}, 2016.

\bibitem{Lustig2014PipeCheckSA}
Daniel Lustig, Michael Pellauer, and Margaret Martonosi.
\newblock Pipecheck: Specifying and verifying microarchitectural enforcement of
  memory consistency models.
\newblock {\em 2014 47th Annual IEEE/ACM International Symposium on
  Microarchitecture}, pages 635--646, 2014.

\bibitem{Manerkar2017RTLCheckVT}
Yatin~A. Manerkar, Daniel Lustig, Margaret Martonosi, and Michael Pellauer.
\newblock Rtlcheck: Verifying the memory consistency of rtl designs.
\newblock {\em 2017 50th Annual IEEE/ACM International Symposium on
  Microarchitecture (MICRO)}, pages 463--476, 2017.

\bibitem{Manerkar2018PipeProofAM}
Yatin~A. Manerkar, Daniel Lustig, Margaret Martonosi, and Aarti Gupta.
\newblock Pipeproof: Automated memory consistency proofs for microarchitectural
  specifications.
\newblock {\em 2018 51st Annual IEEE/ACM International Symposium on
  Microarchitecture (MICRO)}, pages 788--801, 2018.

\bibitem{Trippel2018C}
Caroline Trippel, Daniel Lustig, and Margaret Martonosi.
\newblock Checkmate : Automated exploit program generation for hardware
  security verification.
\newblock 2018.

\bibitem{Manerkar2015CCICheckU}
Yatin~A. Manerkar, Daniel Lustig, Michael Pellauer, and Margaret Martonosi.
\newblock Ccicheck: Using $\mu$hb graphs to verify the coherence-consistency
  interface.
\newblock {\em 2015 48th Annual IEEE/ACM International Symposium on
  Microarchitecture (MICRO)}, pages 26--37, 2015.

\bibitem{Manson2005TheJM}
Jeremy Manson.
\newblock The java memory model.
\newblock In {\em POPL '05}, 2005.

\bibitem{Sewell2010x86TSO}
Peter Sewell, Susmit Sarkar, Scott Owens, Francesco~Zappa Nardelli, and
  Magnus~O. Myreen.
\newblock x86-tso.
\newblock {\em Communications of the ACM}, 53:89 -- 97, 2010.

\bibitem{Batty2011MathematizingCC}
Mark Batty, Scott Owens, Susmit Sarkar, Peter Sewell, and Tjark Weber.
\newblock Mathematizing c++ concurrency.
\newblock In {\em POPL '11}, 2011.

\bibitem{Ahamad2005CausalMD}
Mustaque Ahamad, Gil Neiger, James~E. Burns, Prince Kohli, and Phillip~W.
  Hutto.
\newblock Causal memory: definitions, implementation, and programming.
\newblock {\em Distributed Computing}, 9:37--49, 2005.

\bibitem{Moiseenko2020ReconcilingES}
Evgenii Moiseenko, Anton Podkopaev, Ori Lahav, Orestis Melkonian, and Viktor
  Vafeiadis.
\newblock Reconciling event structures with modern multiprocessors.
\newblock {\em ArXiv}, abs/1911.06567, 2020.

\bibitem{Jeffrey2016OnTA}
Alan Jeffrey and James Riely.
\newblock On thin air reads towards an event structures model of relaxed
  memory.
\newblock {\em 2016 31st Annual ACM/IEEE Symposium on Logic in Computer Science
  (LICS)}, pages 1--9, 2016.

\bibitem{Norris2013CDScheckerCC}
Brian Norris and Brian Demsky.
\newblock Cdschecker: checking concurrent data structures written with c/c++
  atomics.
\newblock {\em Proceedings of the 2013 ACM SIGPLAN international conference on
  Object oriented programming systems languages \& applications}, 2013.

\bibitem{Aronis2018EffectiveTF}
Stavros Aronis.
\newblock Effective techniques for stateless model checking.
\newblock 2018.

\bibitem{Kokologiannakis2020HMCMC}
Michalis Kokologiannakis and Viktor Vafeiadis.
\newblock Hmc: Model checking for hardware memory models.
\newblock {\em Proceedings of the Twenty-Fifth International Conference on
  Architectural Support for Programming Languages and Operating Systems}, 2020.

\bibitem{Berstel1979TransductionsAC}
Jean Berstel.
\newblock Transductions and context-free languages.
\newblock In {\em Teubner Studienb{\"u}cher : Informatik}, 1979.

\bibitem{Sakarovitch2009ElementsOA}
Jacques Sakarovitch.
\newblock Elements of automata theory.
\newblock 2009.

\bibitem{litmus}
Luc Maranget, Jade Alglave, Susmit Sarkar, and Peter Sewell.
\newblock {Litmus: Running Tests against Hardware}.
\newblock In {\em {TACAS'11, 17th International Conference on Tools And
  Algorithms for the Construction and Analysis of Systems}}, Saarbr{\"u}cken,
  Germany, March 2011.

\bibitem{vscale}
LGTMCU.
\newblock vscale.
\newblock \url{https://github.com/LGTMCU/vscale}.
\newblock [Online; accessed 11-05-2021].

\bibitem{tomasulo}
Soham-Das-2021.
\newblock Tomasulo.
\newblock \url{https://github.com/Soham-Das-2021/Tomasulo-Machine}.
\newblock [Online; accessed 11-05-2021].

\bibitem{sdram}
stffrdhrn.
\newblock Sdram controller.
\newblock \url{https://github.com/stffrdhrn/sdram-controller}.
\newblock [Online; accessed 11-05-2021].

\bibitem{Wolf2013YosysAFV}
Clifford Wolf, Johann Glaser, and Johannes Kepler.
\newblock Yosys-a free verilog synthesis suite.
\newblock 2013.

\bibitem{boolector}
Aina Niemetz, Mathias Preiner, and Armin Biere.
\newblock Boolector 2.0.
\newblock {\em J. Satisf. Boolean Model. Comput.}, 9(1):53--58, 2014.

\bibitem{abcTool}
Berkeley~Logic Synthesis and Verification Group.
\newblock Abc: A system for sequential synthesis and verification, release
  70930.
\newblock \url{http://www.eecs.berkeley.edu/~alanmi/abc/}.

\bibitem{Jacob2007MemorySC}
Bruce Jacob, Spencer~W. Ng, and David~T. Wang.
\newblock Memory systems: Cache, dram, disk.
\newblock 2007.

\bibitem{Shasha1988EfficientAC}
Dennis Shasha and Marc Snir.
\newblock Efficient and correct execution of parallel programs that share
  memory.
\newblock {\em ACM Trans. Program. Lang. Syst.}, 10:282--312, 1988.

\bibitem{RVManual}
RISC-V Foundation.
\newblock The risc-v instruction set manual, volume i: User-level isa, document
  version 2.2.

\bibitem{Lustig2016COATCheckVM}
Daniel Lustig, Geet Sethi, Margaret Martonosi, and Abhishek Bhattacharjee.
\newblock Coatcheck: Verifying memory ordering at the hardware-os interface.
\newblock {\em Proceedings of the Twenty-First International Conference on
  Architectural Support for Programming Languages and Operating Systems}, 2016.

\bibitem{Batty2015OverhaulingSA}
Mark Batty, Alastair~F. Donaldson, and John Wickerson.
\newblock Overhauling sc atomics in c11 and opencl.
\newblock {\em Proceedings of the 43rd Annual ACM SIGPLAN-SIGACT Symposium on
  Principles of Programming Languages}, 2015.

\bibitem{Vafeiadis2015CommonCO}
Viktor Vafeiadis, Thibaut Balabonski, Soham~Sundar Chakraborty, Robin Morisset,
  and Francesco~Zappa Nardelli.
\newblock Common compiler optimisations are invalid in the c11 memory model and
  what we can do about it.
\newblock {\em Proceedings of the 42nd Annual ACM SIGPLAN-SIGACT Symposium on
  Principles of Programming Languages}, 2015.

\bibitem{Watt2020RepairingAM}
Conrad Watt, Christopher Pulte, Anton Podkopaev, G.~Barbier, Stephen Dolan,
  Shaked Flur, Jean Pichon-Pharabod, and Shu yu~Guo.
\newblock Repairing and mechanising the javascript relaxed memory model.
\newblock {\em Proceedings of the 41st ACM SIGPLAN Conference on Programming
  Language Design and Implementation}, 2020.

\bibitem{Lahav2017RepairingSC}
Ori Lahav, Viktor Vafeiadis, Jeehoon Kang, Chung-Kil Hur, and Derek Dreyer.
\newblock Repairing sequential consistency in c/c++11.
\newblock {\em Proceedings of the 38th ACM SIGPLAN Conference on Programming
  Language Design and Implementation}, 2017.

\bibitem{Bchi1990OnAD}
J.~Richard B{\"u}chi.
\newblock On a decision method in restricted second order arithmetic.
\newblock 1990.

\bibitem{WolperVardiSistla83}
Pierre Wolper, Moshe~Y. Vardi, and A.~Prasad Sistla.
\newblock Reasoning about infinite computation paths.
\newblock In {\em 24th Annual Symposium on Foundations of Computer Science
  (sfcs 1983)}, pages 185--194, 1983.

\bibitem{Hsiao2021SynthesizingFM}
Yao Hsiao, Dominic~P. Mulligan, Nikos Nikoleris, Gustavo Petri, and Caroline
  Trippel.
\newblock Synthesizing formal models of hardware from rtl for efficient
  verification of memory model implementations.
\newblock {\em MICRO-54: 54th Annual IEEE/ACM International Symposium on
  Microarchitecture}, 2021.

\end{thebibliography}

\onecolumn
\appendix
\section{Supplementary material on theoretical results}

\subsection{Formal model of computation}

In the main section, we described the model of
computation as resembling a transducer with multiple
input tapes (representing instruction streams).
We now provide the formal operational semantics
of this model.

\tikzset{background rectangle/.style={fill=none
}}
\begin{figure}[h]
\centering
\small
\resizebox{\textwidth}{!}{
\begin{tikzpicture}[codeblock/.style={line width=0.5pt, inner xsep=0pt, inner ysep=5pt}]
\node[codeblock] (init) at (current bounding box.north west) {
$
\def\arraystretch{4}
\setlength{\arraycolsep}{5pt}
\begin{array}{cc}
\inferrule{
    \delta = (\qu, (\aninstr_1, \cdots, \aninstr_{|\setcores|}), \qu', \rightact(\acore)) \\
    \aconfiguration = (\yu, \qu, \ve) \\ \enabled{\delta, \aconf} \\ \fst{\ve} \neq \ldash \\\\ 
    \ve(\acore) = \aninstr\cdot w \\ \ve' = \ve[\acore \leftarrow w] \\
    \yu' = \yu[\acore \leftarrow \yu(\acore)\cdot\aninstr]
}{
    \text{Move right}:\quad(\yu, \qu, \ve) \xrightarrow[]{} (\yu', \qu', \ve')
} & \inferrule{
    \delta = (\qu, (\aninstr_1, \cdots, \aninstr_{|\setcores|}), \qu', \copyact(\acore, i, \astage)) \\\\
    1 \leq i \leq |\yu(\acore)| \\ \aconfiguration = (\yu, \qu, \ve) \\ \enabled{\delta, \aconf}
}{
    \text{Execute event}:\quad(\yu, \qu, \ve) \xrightarrow[]{\event{\yu(\acore)[i]}{\astage}} (\yu, \qu', \ve)
} \\[0.1cm]
\inferrule{
    \delta = (\qu, (\aninstr_1, \cdots, \aninstr_{|\setcores|}), \qu', \dropact(\acore, i)) \\
    \aconfiguration = (\yu, \qu, \ve) \\ \enabled{\delta, \aconf} \\\\
    \yu(\acore)[0\cdots i-1] = x \\ \yu(\acore)[i+1 \cdots |U|-1] = y \\ \yu'[\acore \leftarrow x\cdot y]
}{
    \text{Drop}:\quad    (\yu, \qu, \ve) \xrightarrow[]{} (\yu', \qu', \ve)
} &
\inferrule{
    \delta = (\qu, (\aninstr_1, \cdots, \aninstr_{|\setcores|}), \qu', \stayact) \\\\
    \aconfiguration = (\yu, \qu, \ve) \\ \enabled{\delta, \aconf}
}{
    \text{Silent}:\quad (\yu, \qu, \ve) \xrightarrow[]{} (\yu, \qu', \ve)
}
\end{array}$
};
\end{tikzpicture}
}
\caption{Machine transition rules.}
\label{fig:machinesteps}
\end{figure}

Figure \ref{fig:machinesteps} above provides the set of steps that can be taken by the machine.
Recall that each configuration is a triple $(\yu, \qu, \ve)$
where $\yu$ ($\ve$) map each core $\acore$ to the input tape contents
for $\acore$ to the left (right) of the tape head and 
$\qu$ is the control state.
We now describe the transition actions that can be performed by the machine.
The action
$\rightact(\acore)$ moves the tape head for $\acore$ to the right;
$\stayact$ is a silent transition that only changes the control state;
$\copyact(\acore, i, \astage)$ executes the $\astage$-stage event for the instruction at 
$\yu(\acore)[i]$ and $\dropact(\acore, i)$ removes the instruction at $\yu(\acore)[i]$.
We use a 1-based array indexing notation for $\yu(\acore)$.

\subsection{Determing refinability of universal axiomatic semantics}

We give a proof of Lemma \ref{lem:checkref}. Intuitively, this proof 
shows that if an axiomatic semantics does not satisfy refinability, then there is 
a computable bound $b$ such that there exists a program of size atmost $b$
such that some valid execution graph $G$ of $\aprogram$ is a witness to non-refinability, 
i.e. some linearization of $G$ violates the axioms. Then, since the number 
of programs of size $b$ or less are bounded, we get a procedure to determine refinability.

\begin{proof}
    Consider a universal axiomaitc semantics $\anaxiomaticsem$, a program 
    $\aprogram$ and a graph $G$, s.t. $G \models_\aprogram \anaxiomaticsem$ 
    but that does not satisfy refinability, i.e. there is some linearization 
    $G'$ s.t $G \sqsubseteq G'$ and $G' \not\models_\aprogram \anaxiomaticsem$. 
    We can show that there exists a ``small'' subgraph of $G$, 
    say $G_1$ and a linearization, $G_1 \sqsubseteq G_1'$
    such that $G_1 \models_{\aprogram_1} \anaxiomaticsem$ but 
    $G_1' \not\models_{\aprogram_1} \anaxiomaticsem$ for a subprogram $\aprogram_1$
    of $\aprogram$.

    Consider an axiom $\anaxiom = \forall \asymbinstr_1 \cdots \forall \asymbinstr_k ~\phi(\asymbinstr_1, \cdots, \asymbinstr_k)$ 
    from $\anaxiomaticsem$ that the graph $G'$ violates. By the semantics of validity, 
    there must be set of $k$ instructions $\setinstr = \{\aninstr_1, \cdots, \aninstr_k\}$ 
    such that $\phi(\aninstr_1, \cdots, \aninstr_k)$ does not hold in $G'$. 
    Let the program comprised of these instructions instructions 
    (maintaining the cores and the $<_r$ order) be $\aprogram_1$.
    Now we can take the projection of $G$ and $G'$ on the nodes 
    $\{\nodeOf{\aninstr_j.\astage} ~|~ j \in [k], \astage\in\setstages\}$ 
    (and taking a transitive closure on the edges), to get the graphs $G_1$ and $G_1'$, 
    which are executions of the program $\aprogram_1$.

    We have that any atoms over the instructions $\setinstr$
    holds in $G$ iff it holds in $G_1$ and similarly, it holds in $G'$ iff it holds in $G_1'$.
    This follows since we maintain the $<_r$ orderings between the instructions from 
    $\setinstr$ in program $\aprogram_1$.
    Hence, we have that 
    $G_1 \models_{\aprogram_1} \anaxiom$ but $G_1' \not\models_{\aprogram_1} \anaxiom$.

    Let the $K$ be the largest arity of all axioms from $\anaxiomaticsem$.
    This gives us the desired result: if the semantics is not refinable, there exists a ``small''
    program (of size atmost $K$)  which is a witness to non-refinability.
    Then it suffices to search over all programs of size $K$ for a non-refinability witness.
    These programs are finitely many and hence, the procedure terminates.
\end{proof}

\subsection{Infeasibility of finite state synthesis}

We give a proof of Theorem \ref{thm:infeasible}.
\begin{proof}
	Consider a single core program $\aprogram$, with the instruction stream 
	$\istream=\aninstr_0, \cdots, \aninstr_{m-1}$. 
	Now consider any permutation $\pi$ of $\{0, \cdots, m-1\}$. Then, the execution represented by 
	$\aninstr_{\pi(0)}.\stageS, \aninstr_{\pi(1)}.\stageS, \cdots, \aninstr_{\pi(m-1)}.\stageS, 
	\aninstr_{\pi(0)}.\stageT, \aninstr_{\pi(1)}.\stageT, \cdots, \aninstr_{\pi(m-1)}.\stageT$ is a valid 
	execution of the program, and by completeness, is also generated by the transition system. 
	Let the configuration reached by the transition system after generating the events
	$\aninstr_{\pi(1)}.\stageS, \aninstr_{\pi(2)}.\stageS, \cdots, \aninstr_{\pi(n)}.\stageS$ be 
	$(\yu_\pi, \astate_\pi, \ve_\pi)$.
	Now suppose that $|\setstates| < \mathcal{O}(2^m/m)$. Then by the pigeonhole principle,
	there are two permutations $\pi_1 \neq \pi_2$ such that 
	$(\yu_{\pi_1}, \astate_{\pi_1}, \ve_{\pi_1}) = (\yu_{\pi_2}, \astate_{\pi_2}, \ve_{\pi_2})$.
	This holds since the head has only $m$ possible locations on the input tape.
	Since, $\aninstr_{\pi_1(1)}.\stageT, \cdots, \aninstr_{\pi_1(n)}.\stageT$ is accepted from
	$(\yu_{\pi_1}, \astate_{\pi_1}, \ve_{\pi_1})$, 
	we get that $\aninstr_{\pi_1(1)}.\stageS, \cdots, \aninstr_{\pi_1(n)}.\stageS, 
	\aninstr_{\pi_2(1)}.\stageT, \cdots, \aninstr_{\pi_2(n)}.\stageT$ is accepted by the transition system.
	However this execution does not satisfy \lstinline[style=axiom]{ax1} since $\pi_1 \neq \pi_2$ and hence
	the ordering of $\stageS$ and $\stageT$ events must be reversed for some pair of instructions.
\end{proof}

Corollary \ref{corr:imposs} follows since we require an operational model with a (fixed) finite state space
that is sound and complete for all programs.
\section{Supplementary material for operationalization}

\newcommand{\alanguage}{\mathcal{L}}

\subsection{Axiom automata}

We first discuss Lemma \ref{lem:aauto}. This lemma follows 
since $\setsymbevent(\anaxiom)$ and hence
the language (the set of words in the language) is finite.
We give a proof for completeness' sake.

\begin{proof}
We show that if a word $w \in \setsymbevent(\anaxiom)^\perm$ and 
$s(w) \models \phi(\asymbinstr_1, \cdots, \asymbinstr_k)[s]$
for some assignment $s$ that agrees with context $\acontext$, then 
$s'(w) \models \phi(\asymbinstr_1, \cdots, \asymbinstr_k)[s']$
holds for all assignments $s'$ that agree with $\acontext$.

We claim that $s(w) \models \psi[s] \iff s'(w) \models \psi[s']$ 
whenever $s, s'$ have the same context w.r.t. $\phi$ for all sub-formulae $\psi$ of $\phi$.
We can show by structural induction over $\phi$. 
The base case follows since (1) for non-$\hbrel$ atoms, $s, s'$ having the same 
context means that their interpretation is identical and (2) for $\hbrel$ atoms, 
we have $\event{s(\asymbinstr_a)}{\astage_a} \hbarrow \event{s(\asymbinstr_b)}{\astage_b}$
in $s(w)$ whenever we have $\event{s'(\asymbinstr_a)}{\astage_a} \hbarrow \event{s'(\asymbinstr_b)}{\astage_b}$
in $s'(w)$. The inductive case follows trivially.

Hence, we simply need to determine words from $\setsymbevent(\anaxiom)^\perm$
s.t. $s(w) \models \phi[s]$ for some particular $s$ which agrees with $\acontext$.
But this follows since $|\setsymbevent(\anaxiom)|$ is finite, and we can check these 
words one at a time.
\end{proof}

\subsection{Checking extensibility}

Now we prove an intermediate lemma that implies Lemma \ref{lem:checkext}.
Given a (universal) axiom $\anaxiom \equiv \forall 
\asymbinstr_1 \cdots \forall \asymbinstr_k, \phi(\asymbinstr_1, \cdots, \asymbinstr_k)$,
and a context for $\anaxiom$, $\acontext$. We say that a subset
$\setsymbinstr' \subseteq \setsymbinstr(\anaxiom)$
is prefix closed if whenever $\asymbinstr_j \in \setsymbinstr'$ and atom
$\anatom = \asymbinstr_i <_r \asymbinstr_j$ in $\anaxiom$ with $\acontext(\anatom) = \ttrue$ 
then $\asymbinstr_i \in \setsymbinstr'$.

\begin{lemma}
    The following are equivalent for a universal axiomatic semantics $\anaxiomaticsem$
    \begin{itemize}
        \item $\anaxiomaticsem$ satisfies extensibility.
        \item For every axiom $\anaxiom$ in $\anaxiomaticsem$,
        and consistent context $\acontext$, for all prefix closed $\emptyset \subset \setsymbinstr' \subset \setsymbinstr(\anaxiom)$,
        the language of $\axiomautoof{\anaxiom, \acontext}$, $\alanguage$, satisfies
        $$\{\event{\asymbinstr}{\astage} ~|~ \asymbinstr\in\setsymbinstr', \astage \in \setstages\}^\perm\cdot
        \{\event{\asymbinstr}{\astage} ~|~ \asymbinstr\in\setsymbinstr(\anaxiom)\setminus\setsymbinstr', \astage \in \setstages\}^\perm \subseteq \alanguage$$
    \end{itemize}
    \label{lem:autoext}
\end{lemma}

It is easy to see how this proves Lemma \ref{lem:checkext}: 
since the set of contexts and prefix closed subsets is finite 
one can brute force check for inclusion.
We now prove Lemma \ref{lem:autoext}.

\begin{proof}

\noindent\textbf{$\Longrightarrow$} 
Assume that the universal axiomatic semantics $\anaxiomaticsem$ is extensible.
Now, consider an axiom $\anaxiom \in \anaxiomaticsem$ which has arity $k$
and let $\setsymbinstr(\anaxiom) = \{\asymbinstr_1, \cdots, \asymbinstr_k\}$.
Now consider any consistent context $\acontext$. Since the context is consistent, 
there exists an assignment 
$s : \setsymbinstr(\anaxiom) \rightarrow \setinstr$ that agrees with $\acontext$.
Now consider the program $\aprogram$ consisting of instruction streams from 
$\{s(\asymbinstr) ~|~ \asymbinstr \in \setsymbinstr(\anaxiom)\}$ 
(by preserving the $<_r$ order for instructions with same $\coreof{\cdot}$ component).
This program has $|\instrsetof{\aprogram}| = k$ instructions.

Now consider any nonempty, proper subset $\phi \subset \setsymbinstr' \subset \setsymbinstr(\anaxiom)$
that is prefix closed w.r.t $\acontext$ and consider the set of instructions from $\aprogram$ 
corresponding to $\setsymbinstr'$ under the assignment $s$. Since $\setsymbinstr'$ is prefix closed, 
so program corresponding to these instructions is a prefix program $\aprogram'$ of $\aprogram$,
but with $0 < |\instrsetof{\aprogram'}| < k$ (since $\setsymbinstr'$ was a non-empty proper subset).
Consequently the residual program $\aprogram'' = \aprogram \oslash \aprogram'$ also satisfies 
$0 < |\instrsetof{\aprogram''}| < k$.
By the semantics of quantification, (we need distinct assignments), any executions for program $\aprogram'$
and $\aprogram''$ are valid w.r.t $\anaxiom$ and by refinability, any linearized executions are also valid.
This implies that all words in 
$\{\event{\asymbinstr}{\astage} ~|~ \asymbinstr\in\setsymbinstr', \astage \in \setstages\}^\perm\cdot
\{\event{\asymbinstr}{\astage} ~|~ \asymbinstr\in\setsymbinstr(\anaxiom)\setminus\setsymbinstr', \astage \in \setstages\}^\perm$
represent valid executions and must be contained in $\alanguage$ (accepted by $\axiomautoof{\anaxiom, \acontext}$).

\noindent\textbf{$\Longleftarrow$} Now assume that the second condition holds. Consider a program $\aprogram$
and a prefix, residual: $\aprogram', \aprogram'' = \aprogram \oslash \aprogram'$ respectively.
Consider any valid executions of $G', G''$ of $\aprogram', \aprogram''$ w.r.t. $\anaxiom$.
Now consider the graph $G = G' \triangleright G''$ some linearized execution $\atrace$ of $G$.
Now suppose that $G$ is not valid w.r.t. $\anaxiom$ and 
instructions $\aninstr_1, \aninstr_2, \cdots, \aninstr_k$ form a violating assignment to $\phi$.
If all of $I = \{\aninstr_1, \cdots, \aninstr_k\}$ are from $\aprogram$ or $\aprogram'$, this is contradiction
to either $G' \models_{\aprogram'} \anaxiom$ or $G'' \models_{\aprogram''} \anaxiom$ respectively. 
There is a proper subset $\phi \subset I' \subset I$ s.t. instructions from $I'$ are from $\aprogram'$
and those in $I'' = I\setminus I'$ are from $\aprogram''$.
Now consider context $\acontext$ corresponding to the assignment $s = \{\asymbinstr_i \rightarrow \aninstr_i\}$.
Clearly this is a consistent context and hence by the assumption, we have 
$$\{\event{\asymbinstr}{\astage} ~|~ \asymbinstr\in\setsymbinstr', \astage \in \setstages\}^\perm\cdot
        \{\event{\asymbinstr}{\astage} ~|~ \asymbinstr\in\setsymbinstr(\anaxiom)\setminus\setsymbinstr', \astage \in \setstages\}^\perm \subseteq \alanguage$$
for the language $\alanguage$ of $\axiomautoof{\anaxiom, \acontext}$. 
Now the linearized trace $\atrace$ projected to events in $\{\event{\aninstr}{\astage} ~|~ \aninstr \in I, \astage \in \setstages\}$
belongs to the set $$\{s(w) ~|~ w \in \event{\asymbinstr}{\astage} ~|~ \asymbinstr\in\setsymbinstr', \astage \in \setstages\}^\perm\cdot
\{\event{\asymbinstr}{\astage} ~|~ \asymbinstr\in\setsymbinstr(\anaxiom)\setminus\setsymbinstr', \astage \in \setstages\}^\perm\}.$$
This follows since all events of $I'$ preceed those of $I''$ in $\atrace$.
Hence the concrete automaton $\axiomautoof{\anaxiom, s}$ on the (linearized) trace $\atrace$ accepts it, 
in contradiction to the assumption that instructions from $I$ constituted a violation of $\anaxiom$.
\end{proof}

Since the check mentioned in the second condition can be performed effectively, 
this also gives Lemma \ref{lem:checkext}.

\subsection{Proving Theorem 2}

We begin by showing the lemmata mentioned in the main text which we then use to prove Theorem \ref{thm:tcompleteness}.
Before that we recap some auxilliary notation developed in the main text,
and introduce some new notation.

Fix a trace $\atrace$ and consider a trace index $0 \leq j \leq |\atrace|$.
Recall that we defined $\cmall(j)$ and $\nfall(j)$
to be the set of instructions which have 
completed execution of all events and 
not executed any events at step $j$
of the trace respectively.
Based on this, we defined the (maximal)
prefix-closed subset (w.r.t $\leq_r$)
of $\cmall(j)$ as $\cmset(j)$ and the (maximal)
postfix-closed subset of $\nfall(j)$
as $\nfset(j)$.

\[
\begin{array}{rl}
    \cmset(j) &= \{\aninstr ~|~ \forall \aninstr'.~ \aninstr' \leq_r \aninstr \implies \aninstr' \in \cmall(j)\} \\
    \nfset(j) &= \{\aninstr ~|~ \forall \aninstr'.~ \aninstr \leq_r \aninstr' \implies \aninstr' \in \nfall(j)\} \\
    \ipset(j) &= \instrsetof{\aprogram} \setminus (\cmset(j) \cup \nfset(j))
\end{array}    
\]

We define 
(1) the commit-point of the trace
at step $j$ for core $\acore$, denoted as
$\cmtpoint(j, \acore)$
and 
(2) the fetch-point of the trace 
at step $j$ for core $\acore$ denoted as
$\fetpoint(j, \acore)$as follows.
\[
\begin{array}{rl}
    \footnotesize
    \cmtpoint(j, \acore) &= \min\{\lambda(\aninstr) ~|~ \coreof{\aninstr} = \acore \land \aninstr \not\in \cmall(j)\} \\
    \fetpoint(j, \acore) &= \max\{\lambda(\aninstr) ~|~ \coreof{\aninstr} = \acore \land \aninstr \not\in \nfall(j)\}
\end{array}
\]

Intuitively, $\cmtpoint(j, \acore)$ identifies the
last (largest) instruction index for instructions in 
$\cmset(j)$ for each core and $\fetpoint(j, \acore)$
identifies the first (smallest) instruction
index for instructions in $\nfset(j)$ for each core.
The following figure, an expanded version of Figure 
\ref{fig:pipeline} illustrates this notation.

\begin{figure}[h]
    \centering
    \includegraphics[scale=0.4]{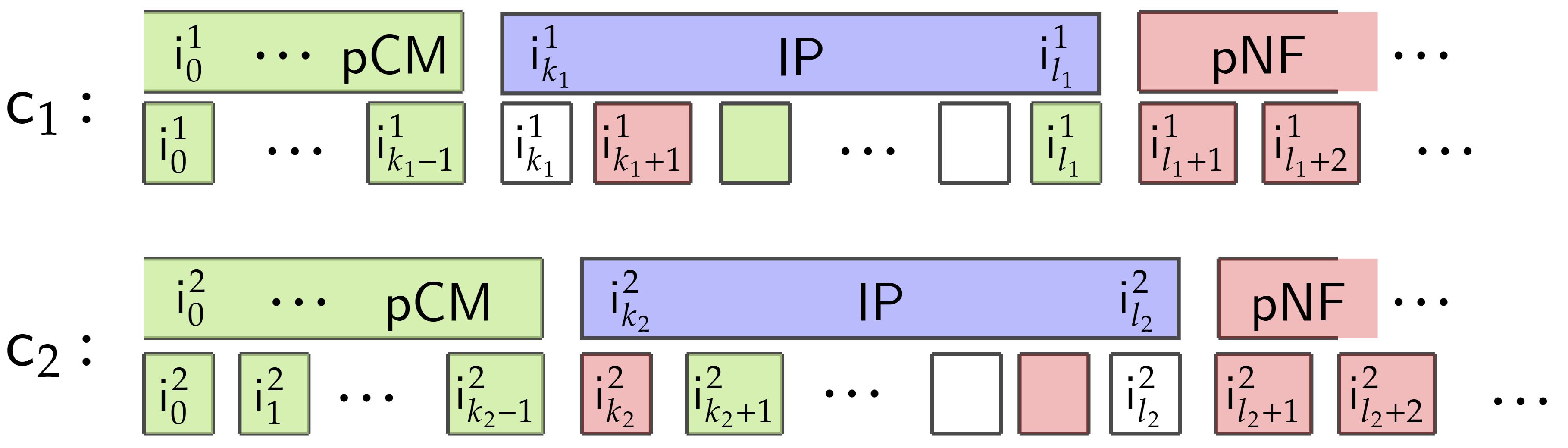}
    \caption{Completed prefix ($\cmset$), in-progess ($\ipset$) and not-fetched postfix ($\nfset$) of
    instructions during execution. We have $\cmtpoint(j, \acore_1/\acore_2) = k_1/k_2$ 
    and $\fetpoint(j, \acore_1/\acore_2) = l_1/l_2$.
    The second row of instructions for each core denote completed
    ($\cmall$) instructions in green, not-fetched ($\nfall$)
    instructions in red and others as uncoloured. Observe that $\cmset$ and $\nfset$ are the maximal pre/post-fixes of $\cmall$ and $\nfall$. As such, former are defined in terms of the latter.
    }
    \label{fig:pipeline-app}
    \vspace{-0.3cm}
\end{figure}

Having defined the auxilliary terminology, we begin by proving 
Lemma \ref{lem:ipbounded}, which states that the 
size of $\ipset(j)$ is bounded.


\begin{proof}
    Consider a $t$ reordering bounded trace $\atrace \in \eventsOf{\aprogram}^\perm$.
    First note that $|\ipset(j)| = \sum_{\acore\in\setcores} (\fetpoint(j, \acore) - \cmtpoint(j, \acore) + 1)$.
    We show that $(\fetpoint(j, \acore) - \cmtpoint(j, \acore) + 1) \leq t$ for all cores $\acore$.

    Now suppose that this quantity exceeded $t$ for some core $\acore$. 
    Let the instructions at labels $\cmtpoint(j, \acore)$ 
    and $\fetpoint(j, \acore)$ be $\aninstr_1$ and $\aninstr_2$ respectively.
    By definition of $\cmtpoint(\cdot)$ and $\fetpoint(\cdot)$ we know that 
    $\aninstr_1$ has not yet executed some event at $j$, say $\event{\aninstr_1}{\astage-1}$ 
    and $\aninstr_2$ has executed atleast one event till $j$, say $\event{\aninstr_2}{\astage_2}$.
    However, we have $\diffr(\aninstr_1, \aninstr_2) = \fetpoint(j, \acore) - \cmtpoint(j, \acore) \geq t$ 
    contradicting the $t$-reordering boundedness assumption.
\end{proof}

Before moving to the proofs of Lemma \ref{lem:activeinst}, 
we give an example that illustrates $\idxstart, \idxend, \idxpfxend$.
\begin{example}
    Consider a program in which 
    the instruction stream is $\istream = \aninstr_0 \cdot \aninstr_1 \cdot \aninstr_2$.
    Let the stages be $\setstages = \{\stageS, \stageT\}$.
    Consider the following trace $\atrace$.
\[\arraycolsep=0.2cm
\begin{array}{c c c c c c c}
    \atrace   & \event{\aninstr_0}{\stageS} & \event{\aninstr_2}{\stageS} & \event{\aninstr_0}{\stageT}
                & \event{\aninstr_1}{\stageS} & \event{\aninstr_2}{\stageT} & \event{\aninstr_1}{\stageT} \\
    \text{indices} & 1 & 2 & 3 & 4 & 5 & 6
\end{array}
\]
Now we have $\idxstart(\aninstr_0) = 1$ and $\idxend(\aninstr_0) = \idxpfxend(\aninstr_0) = 3$ as usual
(since the only instruction $\leq_r$ w.r.t. $\aninstr_0$ is itself). Similarly we have 
$\idxstart(\aninstr_1) = 4$ and $\idxend(\aninstr_0) = \idxpfxend(\aninstr_0) = 6$.
We have $\idxstart(\aninstr_2) = 2$.
However $\idxend(\aninstr_2) \neq \idxpfxend(\aninstr_2)$, since while $\idxend(\aninstr_2) = 5$,
not all instructions ($\aninstr_1$) that are $\leq_r \aninstr_2$ have concluded by index 5.
Hence, $\idxpfxend(\aninstr_2) = 6$.
\end{example}

Now, we prove an equivalent characterization of the second condition for 
$t$-reordering boundedness.
We say that $\aninstr_1 \Bowtie \aninstr_2 \cdots \Bowtie \aninstr_k$, in words 
that these instructions form a \textit{coupling chain}, 
if we have $\coupled(\aninstr_i, \aninstr_{i+1})$
for each $1 \leq i < k$.
The length of such a coupling chain is considered to be $k$.
Note that two instructions are $k$-coupled if they are the endpoints 
of some coupling chain of length $k$.

\begin{lemma}
    \label{lem:treorder-gen}
For any $k \in \nat$ there exists a bound $t_k \in \nat$ 
(independent of program $\aprogram$ and axiomatic model) such that:
for any any $t$-reordering bounded trace $\atrace$ of program $\aprogram$, and
for any instructions $\aninstr, \aninstr'$ on the same core, if there exists a coupling chain
$\aninstr \Bowtie \aninstr_1 \Bowtie \aninstr_2 \cdots  \aninstr_k \Bowtie \aninstr'$ such that
then $|\diffr(\aninstr, \aninstr')| < t_k$.
\end{lemma}

The original definition of $t$-reordering boundness states the above for 
coupling $k = 3$, the above lemma generalizes to $k$.
Intuitively, this follows by an analysis of what it means for instructions to overlap in a trace.
\begin{proof}
    Consider some $t$-reordering bounded trace $\atrace$ for program $\aprogram$, 
    and two instructions $\aninstr, \aninstr'$ on the same core, say $\acore$
     and suppose there is coupling chain 
    $\aninstr \Bowtie \aninstr_1 \Bowtie \cdots \Bowtie \aninstr_k \Bowtie \aninstr'$.
    Now (since the $\coupled(\cdot, \cdot)$ relation is symmteric) WLOG, 
    let $\aninstr <_r \aninstr'$ in the program. Now we have two cases.
    
    \noindent\textbf{Case 1} Some event of $\aninstr'$ executes before an event of $\aninstr$: in this case, we have that 
        $\diffr(\aninstr, \aninstr') < t$ by the first condition of $t$-reordering boundedness
        
    \noindent\textbf{Case 2} All events of $\aninstr'$ execute strictly after those of $\aninstr$: in this case, let 
        $\aninstr^\acore_1, \aninstr^\acore_2, \cdots, \aninstr^\acore_l$ be the instructions from
        $\acore$ such that some event of each of $\aninstr^\acore_i$ has executed between the indices
        $\idxend(\aninstr)$ and $\idxstart(\aninstr')$ (last event of $\aninstr$ and first event of $\aninstr'$). 
        
        We claim for each $\aninstr^\acore_i$
        there exists an instruction from $\aninstr_j \in \{\aninstr_1, \cdots, \aninstr_k\}$ 
        such that $\coupled(\aninstr^\acore_i, \aninstr_j)$. This is easy to see: if some event of $\aninstr^\acore_i$
        happens between the executions of $\aninstr, \aninstr'$ and this entire interval is spanned by an overlapping
        chain of intervals $\aninstr_1 \Bowtie \cdots \Bowtie \aninstr_k$ then the interval 
        $[\idxstart(\aninstr^\acore_i), \idxpfxend(\aninstr^\acore_i)]$ must overlap with one of these.
        must have happened within the execution interval of some $\aninstr_j$.
        Now by the second condition of $t$-reordering boundedness, atmost $t$ instructions on core $\acore$
        can be coupled with any of the $\aninstr_j$s. Consequently we have $l < k\cdot t$.

        Now we will show that almost all of instructions lying between $\aninstr, \aninstr'$ w.r.t. the $<_r$
        order must be one of $\aninstr^\acore_1, \cdots, \aninstr^\acore_l$
        Using the first $t$-reordering boundedness condition, 
        we must have that atmost $t-1$ of the instructions \textit{after} $\aninstr$
        (in $<_r$ order) can execute \textit{before} $\aninstr$ 
        and similarly atmost $t-1$ of the instructions \textit{before} $\aninstr'$
        (in $<_r$ order) can execute \textit{after} $\aninstr'$. 
        Hence, we have that $\diffr(\aninstr, \aninstr') \leq l + 2(t-1) < t(k+2)-2$.
        This gives us the (axiom, program independent) bound of $t_k = t(k+2)-2$.
    Since the value of $t$ from the first case is smaller than $t(k+2)-2$ for all $k$, $t_k = t(k+2)-2$
    the bound of $t_k = t(k+2)-2$ suffices.
\end{proof}

Now we are in a position to prove Lemma \ref{lem:activeinst}.
Intuitively, we will make use of the generalization result from Lemma \ref{lem:treorder-gen},
to show that on each core there can be only so many instructions which are $k$ coupled with
a given instruction from $\ipset(j)$. Then by Lemma \ref{lem:ipbounded}, since the 
size of $|\ipset(j)|$ itself is bounded we get the desired result.

\begin{proof}{(Lemma \ref{lem:activeinst})}
    To start off, we note that if $\aninstr_1$ and $\aninstr_2$ are $k$-coupled and
    if $\aninstr_2$ and $\aninstr_3$ are $k$-coupled then $\aninstr_1$ and $\aninstr_3$
    are $(2k-1)$-coupled. 

    Now consider an instruction $\aninstr \in \ipset(j)$. Then on any core $\acore$, there can only 
    exist $t_{2k-3}$ instructions with which $\aninstr$ is $k$-coupled. Suppose to the contrary there
    were more than that. Then there are two instructions $\aninstr_1, \aninstr_2$ on $\acore$
    such that $|\diffr(\aninstr_1, \aninstr_2)| \leq t_{2k-3}$ which are both $k$-coupled with $\aninstr$.
    Then we know by the earlier observation that $\aninstr_1$ and $\aninstr_2$ are $(2k-1)$-coupled
    which by Lemma \ref{lem:treorder-gen} implies that $|\diffr(\aninstr_1, \aninstr_2)| < t_{2k-3}$
    contradicting the assumption.

    Then, across all cores and all instructions in $\ipset(j)$
    there are atmost $|\ipset(j)|\cdot n \cdot t_{2k-3}$ instructions in $\cmset(j) \cup \ipset(j)$
    that are $k$-coupled with some instruction in $\ipset(j)$. Hence the lemma holds
    for $b_k = |\ipset(j)|\cdot n \cdot t_{2k-3}$ (which is independent of the program and axiom).
\end{proof}

\newcommand{\Kcoup}{\mathsf{AC}_K}

Now we go from active instructions to active automata. 
For this we fix an axiom $\anaxiom$ of arity $K$.
Intuitively, active instructions represent instructions for which
we must maintain the event ordering (even if the instruction is committed).
A (concrete) active automata is an automata over only active instructions.
We prove that for an assignment $s : \setsymbinstr \rightarrow \cmset(j) \cup \ipset(j)$, 
if any of the instructions in $s(\setsymbinstr(\anaxiom))$ are not in 
$\Kcoup(j)$ then the automaton $\axiomautoof{\anaxiom, s}$ need not be explicitly
maintain since it is guaranteed to accept. We make use of extensibility 
via its equivalent characterization in Lemma \ref{lem:autoext}.
We define \textit{active} axiom automata at trace index $j$.

\paragraph{Active axiom automata}

Once again consider a trace $\atrace = \anevent_1 \cdot \anevent_2 \cdot \anevent_{|\atrace|}$. 
The set of \textit{active} (concrete) axiom automata at trace index $0 \leq j \leq |\atrace|$, denoted as 
$\activeauto_{\anaxiom, \aprogram}(j)$ is a subset of $\aautoset(\anaxiom, \cmset(j)\cup\ipset(j))$
defined as follows. Let $E(j, s) = \{\anevent_1, \cdots, \anevent_j\} \cap s(\setsymbevent(\anaxiom))$
be the events that the automaton $\axiomautoof{\anaxiom, s}$ has transitioned on till $j$.
The automaton $\axiomautoof{\anaxiom, s}$ is said to be \textit{active} 
at $j$ if there is a path from the current state to a non-accepting state on some word from 
$(s(\setsymbevent(\anaxiom)) \setminus E(j, s))^\perm$. Intuitively, this means
that the automaton can still reject the trace $\atrace$ by consuming the remainder of the trace.
We relate coupling chains with activeness of automata.

We define $k$-coupling by a set of instructions. Given a trace $\atrace$ and a set of instructions
$I$, we say that two instructions $\aninstr, \aninstr' \in I$
are $k$-coupled by $I$ if there exist $\aninstr_1, \cdots, \aninstr_{k-2} \in I$ such that
$\aninstr \Bowtie \aninstr_1 \Bowtie \cdots \Bowtie \aninstr_{k-2} \Bowtie \aninstr'$
form a coupling chain (of length $k$). The only additional requirement is that the intermediate
instructions be from the prescribed set $I$.

\begin{lemma}
    \label{lem:activeauto}
    Consider a trace $\atrace$ and index $0 \leq j \leq |\atrace|$.
    For some (injective) assignment 
    $s : \setsymbinstr(\anaxiom) \rightarrow \cmset(j) \cup \ipset(j)$
    the automaton $\axiomautoof{\anaxiom, s}$ is active only if 
    each instruction in $I = s(\setsymbinstr(\anaxiom))$ is
    $K$-coupled by $I$ to some instruction in $\ipset(j) \cap I$, 
    where $K$ is the arity of $\anaxiom$.
\end{lemma}
\begin{proof}
    Assume that some instruction $\aninstr \in I$ is not $K$-coupled by $I$ 
    to any instruction in $\ipset(j) \cap I$.
    Then there exists a partitioning of $I = I_a \uplus I_b$ such that 
    $\aninstr \in I_a$, $\ipset(j) \cap I \subseteq I_b$ and 
    there do not exist any $\aninstr' \in I_a$ and $\aninstr'' \in I_b$
    such that $\coupled(\aninstr', \aninstr'')$.
    This holds since $|I| = K$ and we can define $I_b$
    to be the $\coupled$-closure starting from $\ipset(j) \cap I$.

    Now, by the definition of $\coupled$, $I_a$ is a prefix closed subset of $I$.
    Hence by equivalent characterization of extensiblity in Lemma \ref{lem:autoext}, 
    we know that there is no permutation over the events of $I_b$ such that automaton 
    does not accept. Consequently, the automaton is not active.
\end{proof}

Finally, combining Lemma \ref{lem:activeinst} and Lemma \ref{lem:activeinst},
we get that the number of active automata must be bounded.

\begin{corollary}
    For each universal axiom $\anaxiom$, 
    there is a bound $d \in \nat$ s.t. for any program $\aprogram$ and 
    $t$-reordering bounded trace $\atrace$ of $\aprogram$,
    we have $|\activeauto_{\anaxiom, \aprogram}(j)| \leq d$ for all $0 \leq j \leq |\atrace|$.
\end{corollary}

\begin{proof}
    This follows since each active automata must have all its corresponding
    instructions to be $K$-coupled to some instruction in $\ipset(j)$.
    However, by Lemma \ref{lem:activeinst}, the number of instructions
    $K$-coupled to some instruction $\ipset(j)$ is bounded above by $b_K$
    Consequently, the number of active automata is bounded (by the loose bound $\sim b_K^K$).
\end{proof}

\subsection{Compiling automata into a transition system}

\begin{wrapfigure}[15]{r}{0.5\textwidth}
    \vspace{-0.5cm}
    \begin{minipage}{0.5\textwidth}
        \begin{algorithm}[H]                
            \caption{}
        \KwData{$\activeinst{K}$}
        \While{$\exists \acore~ \fst{\ve(\acore)} \neq \ldash$ or 
            some event in $\yu$ is unscheduled}{
            $\anevent \leftarrow$ unscheduled event from $\yu$\;
            \For{$\anaxiom \in \anaxiomaticsem$}{  
                \For{$a \in \axiomautoof{\anaxiom, \activeinst{K}}$}{
                    transition $a$ on $\anevent$\;
                    \lIf(\tcp*[h]{bad}){$a$ rejects}{FAIL}
                }
            }
            \If{$\yu(\acore)[1]$ has completed all events for some $\acore$}{
                $\dropact(\acore, 1)$; $\rightact(\acore)$; update $\activeinst{K}$
            }
        }
        goto $\astate_\final$
        \end{algorithm}
    \end{minipage}
    \caption{Operational loop}
    \label{fig:operationalloop}
\end{wrapfigure}

The Figure \ref{fig:operationalloop} alongside provides an informal pseudocode 
for the operational model for $\anaxiomaticsem$ 
where $K$ is the maximum arity of axioms in $\anaxiomaticsem$.
The machine maintains the instructions from $\ipset$ in the history $\yu$ 
(which is possible for bounded $\yu$ by Lemma \ref{lem:ipbounded} for $h=t$). It maintains
instructions from $\activeinst{K}$ 
in its finite state (possible because of Lemma \ref{lem:activeinst}).
At each loop iteration the machine non-deterministically chooses
an event to schedule. If adding that event moves some automaton
to the rejecting state then the machine moves to an (absorbing) reject state.
Otherwise, the machine checks whether an instruction on some core can be ``committed'' (by the $\dropact$ action); if so a new instruction takes its place (by the $\rightact$ action). Finally it updates $\activeinst{K}$ according to the new set of instructions in $\ipset$.
This repeats until all instructions have been executed.
We have the following.
\begin{claim}
    \emph{\textbf{Operational Loop}} satisfies conditions of Theorem \ref{thm:tcompleteness}.
\end{claim}
In the following section, adopt the above construction when
putting everything together and proving Thm. \ref{thm:tcompleteness}.

\subsection{From axiom automata to an operational transition system}

We now discuss how the earlier boundedness results imply a finite
state bounded history transition system.
Intuitively, we show that we can actually generate the
active axiom automata on the fly using a finite state space and a 
history parameter of $h = t$, i.e.
our history parameter conincides with the reorering bound 
(this is not surprising given that the instructions from $\yu$
are the ones that can be reordered with respect to each other).
Let $K$ be max arity over all axioms in $\anaxiomaticsem$.

We now present a set of invariants for our transition system and then 
define the system itself. 
To that effect consider a trace $\atrace = \anevent_1\cdot\anevent_2\cdots\anevent_{|\atrace|}$ 
and a trace index $0 \leq j\leq |\atrace|$. For all $j$ we have the following invariants:
The first invariant is that 
the history (left of tape head) $\yu$ contains precisely the in-progress 
instructions at each step.
\begin{equation*}
    \text{(1)}\quad \ipset(j, \acore) = \yu_j(\acore)
\end{equation*}
where $\ipset(j, \acore)$ denotes the instructions in $\ipset(j)$ on core $\acore$.
This is feasible by Lemma \ref{lem:ipbounded}, since $|\ipset(j, \acore)| \leq t = h$.
The second set of invariants are regarding the finite state maintained by the system:
\begin{itemize}[leftmargin=1cm]
    \item[(2a)] The finite control state maintains the ordering of
    events for $\Kcoup(j)$, i.e. the set of active instructions at $j$. 
    By Lemma \ref{lem:activeinst} this set is bounded and hence can be represented in
    finite state. (Also note that using this ordering the machine can 
    infer the coupling chains formed between these intructions).
    \item[(2b)] The control state also maintains the valuation of all possible 
    $\predsof{\anaxiom}$ 
    atoms over instructions in $\Kcoup(j)$ (which by definition also contains 
    the instructions from $\ipset(j)$).
    Note that even if the instruction labels are natural numbered (not finite),
    the transition system does not need to maintain the labels but rather only the relative ordering.
    \item[(2c)] Finally the control tracks the events from instructions in $|\ipset(j)|$
    which have been scheduled - once again being possible because of boundedness of $|\ipset(j)|$.
\end{itemize}

We do not go into the intricacies of the exact way in which the 
above information is encoded - the boundedness properties imply that it can.
Now we describe the operational model.
The model exactly follows the \textbf{Operational Loop}
pseudocode presented in \S\ref{sec:operational}.
Intuitively, the model 
has three macro actions: (a) event scheduling, (b) commit-fetch and
(c) recycling axiom automata. At each macro step
the model cycles over these actions (i.e. it performs event scheduling,
followed by commit-fetch and then recycling) following which it returns back to 
the start of the loop for the next step. Each macro step leads to
the scheduling of one event ($j$ increments by 1).
We describe these in turn.

\newcommand{\kmonecoup}{\mathsf{AC}_{K-1}}

\paragraph{Event scheduling}
The operational model has a schedule transition for each
index into the history, $1 \leq i \leq h$ and core $\acore$ and stage $\astage$.
This transition is contingent on the fact that the event has not been scheduled before,
which is tracked by (2c).
Either scheduling the event $\event{\yu(\acore)[i]}{\astage}$
moves one of the automata $\activeauto_{\anaxiom, \aprogram}(j)$ 
for some axiom $\anaxiom$
into the rejecting state, and the operational model moves into the rejecting state
as well, or the operational model updates the control to maintain (2a,2c).
The set of these active automata can be determined since the $\predsof{\anaxiom}$
valuations of all instructions is known (by 2b).
It maintains (2a) by appending the event to the current ordering over $\Kcoup(j)$ 
and (2c) by flagging it as executed.

\paragraph{Commit-fetch}

After scheduling an event the system checks 
whether the instruction at the leftmost position of $\yu(\acore)$ 
has completed scheduling
of all its events (possible because of (2c)). 
If so it is dropped by the action $\dropact(\acore, 1)$ 
from $\yu$ and the subsequent step the tape head for $\acore$ 
is moved one step towards the right. This proves invariant (1).

In the case we add a new instruction to $\ipset(j)$, say $\aninstr$ 
we crucially need to ensure 
that the invariants (2a,2b) are maintained. That is, we need to ensure that 
the ordering and
predicates for instruction in new $\Kcoup$, $Kcoup(j+1)$ can be computed. 
However, this is possible since the new instruction is not directly coupled
with any instruction from $\cmset(j)$ (all those instructions completed 
execution of all events before $\aninstr$ was fetched). 
Hence any coupling chain from $\aninstr$ to $\cmset(j)$
must pass through some instruction in $\ipset(j)$. 
Moreover since we want length $K$ chains from 
$\aninstr$, we need to look for $K-1$ chains from $\ipset(j)$. 
However, we have that $\kmonecoup(j) \subseteq \Kcoup(j)$ and hence any 
instructions from $\cmset(j)$ that are $K$-coupled 
with the new instruction $\aninstr$
are already in the instructions maintained in the finite state space (by 2b)
and these chains can be computed (by 2a).

\paragraph{Recycling old instructions}

After the earlier two macro steps, the machine tries to recycle
old instructions that are no more in $\Kcoup$.
While fetching new instructions on $\acore$ (moveing one step to the right), 
the instruction to the leftmost of $\yu(\acore)$, say $\aninstr'$ 
is committed (by $\dropact$).
Hence, we can remove all the instructions from $\cmset(j)$ that were only $K$-coupled
with $\aninstr'$ and no other instruction, since they will not be in $\cmset(j+1)$.
Lemma \ref{lem:activeinst} shows that $|\Kcoup(j)|$ is always bounded, hence
the model will always be able to free up space allocated to the 
now-inactive instructions from $\cmset(j)$.
By recycling old instructions, we also recycle the axiom automata
over those instructions and maintain the invariants for $j+1$.

\paragraph{Putting everything together}
Thus, at each step, 
the axiomatic model schedules an event from $\ipset(j)$.
On scheduling that event if the earliest (leftmost) instruction on any core
has completed execution of all its events, it performs the fetch-commit 
and recycling steps. As mentioned earlier, all these operations can be 
performed in finite space by the bounding results Lemma 
\ref{lem:ipbounded},\ref{lem:activeinst},\ref{lem:activeauto}.
Finally the system repeats this cycle at each macro step of the execution.
When the system as executed all the events from $\yu$ and there are no more instructions
to be executed it moves to the accepting state.

\paragraph{Soundness and $t$-Completeness}
Firstly it is easy to see that any trace $\atrace$ generated by the above 
system generates is a member of $\eventsOf{\aprogram}^\perm$.
This holds since the machine tracks the events from $\ipset(j)$
that have been scheduled and hence each event is scheduled atmost once.
Each event is also scheduled since the Commit-Fecth requires that all
events of the dropped instruction be complete. Finally the system accepts only when 
there are no more instructions to be fetched (on any core) and all events from
$\yu$ have been scheduled.

Now suppose the model generates an unsound trace $\atrace$
with respect to axiom 
$\forall \asymbinstr_1, \cdots \asymbinstr_k, \phi(\aninstr_1, \cdots, \aninstr_k)$.
Then for some assignment $s$, $\phi[s]$ must have been false.
Let $\aninstr$ be the instruction in $s(\setsymbinstr(\anaxiom))$
that had its event executed last in $\atrace$. Let the trace index of that 
event in $\atrace$ be $j$. Then we have two possibilities:
(1) $\axiomautoof{\anaxiom, s} \in \activeauto_{\anaxiom, \aprogram}(j)$
or (2) $\axiomautoof{\anaxiom, s} \not\in \activeauto_{\anaxiom, \aprogram}(j)$.
In the first case, we know from invariants (2a,2b) and Lemma \ref{lem:aauto} (property of axiom automata) 
that adding the event to the trace would made the machine move into the rejecting state,
contradicting the claim.
In the second case, the definition of axiom automata means that $\axiomautoof{\anaxiom, s}$
and the fact $k \leq K$ means that it
would not have rejected the ordering, once again contradicting the assumption.

On the other hand, $t$-completeness follows since we maintain a per-core history of length $h = t$
and that at each macro-step we non-deterministically schedule one of the events 
corresponding to instructions in $\ipset(j)$. If at some step $j$ 
the model moves into a bad state then that means that the current 
trace already transitioned an axiom automata into a rejecting state. By Lemma \ref{lem:aauto}
(characterization of axiom automata), this means that there was some axiom 
$\forall \asymbinstr_1, \cdots \asymbinstr_k, \phi(\aninstr_1, \cdots, \aninstr_k)$ and assignment
$s$, such that $\phi[s]$ did not hold and hence, the trace was also invalid with respect to the 
axiomatic model. $\hfill\qed$

\end{document}